\documentclass[a4paper,UKenglish,cleveref,autoref,thm-restate]{lipics-v2021}
\pdfoutput=1 

\hideLIPIcs

\usepackage[dvipsnames]{xcolor}

\usepackage{tikz-cd,xspace,mathtools,stmaryrd}

\usepackage[T1]{fontenc}
\DeclareMathAlphabet{\mathbf}{OT1}{jkpx}{sb}{n}
\SetMathAlphabet{\mathbf}{bold}{OT1}{jkpx}{b}{n}
\usetikzlibrary{graphs,decorations.pathmorphing,decorations.markings}

\DeclareFontFamily{U}{mathb}{}
\DeclareFontShape{U}{mathb}{m}{n}{
     <-5.5> mathb5
  <5.5-6.5> mathb6
  <6.5-7.5> mathb7
  <7.5-8.5> mathb8
  <8.5-9.5> mathb9
  <9.5-11>  mathb10
  <11->     mathb12
}{}
\DeclareSymbolFont{mathb}{U}{mathb}{m}{n}
\DeclareMathSymbol{\leftleftharpoons}      {3}{mathb}{"D8}
\DeclareMathSymbol{\rightrightharpoons}    {3}{mathb}{"D9}
\DeclareMathSymbol{\leftbarharpoon}        {3}{mathb}{"DC}
\DeclareMathSymbol{\rightbarharpoon}       {3}{mathb}{"DD}
\DeclareMathSymbol{\barleftharpoon}        {3}{mathb}{"DE}
\DeclareMathSymbol{\barrightharpoon}       {3}{mathb}{"DF}
\DeclareMathSymbol{\leftrightharpoon}      {3}{mathb}{"E0}
\DeclareMathSymbol{\rightleftharpoon}      {3}{mathb}{"E1}

\DeclareFontFamily{U}{mathx}{\hyphenchar\font45}
\DeclareFontShape{U}{mathx}{m}{n}{
      <5> <6> <7> <8> <9> <10>
      <10.95> <12> <14.4> <17.28> <20.74> <24.88>
      mathx10
      }{}
\DeclareSymbolFont{mathx}{U}{mathx}{m}{n}
\DeclareFontSubstitution{U}{mathx}{m}{n}
\DeclareMathAccent{\widecheck}{0}{mathx}{"71}
\DeclareMathAccent{\wideparen}{0}{mathx}{"75}

\theoremstyle{definition}
\theoremstyle{remark}

\colorlet{Green}{OliveGreen}
\colorlet{Red}{WildStrawberry}
\colorlet{Purple}{Plum}
\colorlet{Blue}{Cerulean}
\colorlet{Gray}{black!50!gray}

\definecolor{mygreen}{RGB}{27,158,119}
\definecolor{myred}{RGB}{238,153,170}

\definecolor{myblue}{RGB}{117,112,179}

\renewenvironment{quote}{%
  \list{}{%
    \leftmargin0.3cm   
    \rightmargin\leftmargin
  }
  \item\relax
}
{\endlist}

\newcommand{\ifarg}[2]{\ifthenelse{\equal{#1}{}}{}{#2}}

\newcommand{\tightoverset}[2]{\mathop{#2}\limits^{\vbox to -.5ex{\kern-0.75ex\hbox{$#1$}\vss}}}
\newcommand{\tightunderset}[2]{\mathop{#2}\limits_{\vbox to .5ex{\kern-1.5ex\hbox{$#1$}\vss}}}

\renewcommand{\ll}{\left}
\newcommand{\rr}{\right}
\renewcommand{\b}[1]{\ll[#1\rr]}
\newcommand{\p}[1]{\lp#1\rp}

\renewcommand{\and}{\text{ and }}

\newcommand{\into}{\hookrightarrow}

\newcommand{\lp}{\left(}
\newcommand{\rp}{\right)}

\renewcommand{\ul}[1]{\underline{#1}}

\newcommand{\meet}{\wedge}
\newcommand{\join}{\vee}

\newcommand{\Meet}{\bigwedge}
\renewcommand{\Join}{\bigvee}

\newcommand{\subb}{\sqsubseteq}

\makeatletter
\DeclareRobustCommand{\cev}[1]{%
  \mathpalette\do@cev{#1}%
}
\newcommand{\do@cev}[2]{%
  \fix@cev{#1}{+}%
  \reflectbox{$\m@th#1\vvec{\reflectbox{$\fix@cev{#1}{-}\m@th#1#2\fix@cev{#1}{+}$}}$}%
  \fix@cev{#1}{-}%
}
\newcommand{\fix@cev}[2]{%
  \ifx#1\displaystyle
    \mkern#23mu
  \else
    \ifx#1\textstyle
      \mkern#23mu
    \else
      \ifx#1\scriptstyle
        \mkern#22mu
      \else
        \mkern#22mu
      \fi
    \fi
  \fi
}

\newcommand{\id}{\mathsf{1}}

\newcommand{\R}{\mathbb{R}} 

\newcommand{\A}{\mathcal{A}}
\newcommand{\B}{\mathcal{B}}

\newcommand{\F}{\mathcal{F}}

\newcommand{\I}{\mathcal{I}}

\renewcommand{\H}{\mathrm{H}}

\newcommand{\NN}{\mathbb{N}}

\newcommand{\RR}{\mathbb{R}}

\newcommand{\e}{\varepsilon}

\newcommand{\mathwrap}[1]{\ensuremath{#1}\xspace}

\newcommand{\cat}[1]{\mathwrap{\mathbf{#1}}}

\newcommand{\Cat}{\ensuremath{\mathbf{Cat}}\xspace}
\newcommand{\Set}{\ensuremath{\mathbf{Set}}\xspace}
\newcommand{\Top}{\ensuremath{\mathbf{Top}}\xspace}

\newcommand{\Mch}{\ensuremath{\mathbf{Mch}}\xspace}

\let\vvec\vec

\newcommand{\vect}[1][]{\ensuremath{\mathbf{vec^{#1}}}\xspace}
\renewcommand{\vec}[1][]{\vect[#1]}

\newcommand{\Fac}[1]{\mathwrap{\mathbf{Fac}\ifarg{#1}{(\mathbf{#1})}}}
\newcommand{\Fact}[1]{\Fac{#1}}

\renewcommand{\Bar}[1][]{\ensuremath{\mathbf{Bar_{#1}}}\xspace}

\newcommand{\op}[1]{#1^{\mathsf{op}}}

\newcommand{\vor}{\mathrm{Vor}}
\newcommand{\del}{\mathrm{Del}}

\renewcommand{\lim}{\mathrm{lim}}

\newcommand{\Sub}{\mathrm{Sub}}

\renewcommand{\bar}{\mathsf{Bar}}

\newcommand{\card}{\mathsf{card}}
\newcommand{\bary}{\mathsf{bary}}

\newcommand{\im}{\mathrm{im}\,}

\newcommand{\rk}{\mathrm{rk}\,}
\newcommand{\rank}{\mathrm{rank}\,}

\newcommand{\dist}{\mathrm{d}}

\makeatletter
\providecommand*{\twoheadrightarrowfill@}{\arrowfill@\relbar\relbar\twoheadrightarrow}
\providecommand*{\twoheadleftarrowfill@}{\arrowfill@\twoheadleftarrow\relbar\relbar}
\providecommand*{\xtwoheadrightarrow}[2][]{\ext@arrow 0579\twoheadrightarrowfill@{#1}{#2}}
\makeatother

\providecommand*{\xtwoheadleftarrow}[2][]{\ext@arrow 5097\twoheadleftarrowfill@{#1}{#2}}

\newcommand{\xhooktwoheadrightarrow}[2][]{%
  \lhook\joinrel
  \ext@arrow 0359\rightarrowfill@ {#1}{#2}%
  \mathrel{\mspace{-15mu}}\rightarrow}

\newcommand*{\xtworightarrow}[2]{\mathrel{
  \settowidth{\@tempdima}{$\scriptstyle#1$}
  \settowidth{\@tempdimb}{$\scriptstyle#2$}
  \ifdim\@tempdimb>\@tempdima \@tempdima=\@tempdimb\fi
  \mathop{\vcenter{
    \offinterlineskip\ialign{\hbox to\dimexpr\@tempdima+1em{##}\cr
    \rightarrowfill\cr\noalign{\kern.5ex}
    \rightarrowfill\cr}}}\limits^{\!#1}_{\!#2}}}

\newcommand*{\xtwolongrightarrow}[2]{\mathrel{
  \settowidth{\@tempdima}{$\scriptstyle#1$}
  \settowidth{\@tempdimb}{$\scriptstyle#2$}
  \ifdim\@tempdimb>\@tempdima \@tempdima=\@tempdimb\fi
  \mathop{\vcenter{
    \offinterlineskip\ialign{\hbox to\dimexpr\@tempdima+1em{##}\cr
    \longrightarrowfill\cr\noalign{\kern.5ex}
    \longrightarrowfill\cr}}}\limits^{\!#1}_{\!#2}}}

\newcommand*{\xthreerightarrow}[1]{\mathrel{
  \settowidth{\@tempdima}{$\scriptstyle#1$}
  \mathop{\vcenter{
    \offinterlineskip\ialign{\hbox to\dimexpr\@tempdima+1em{##}\cr
    \rightarrowfill\cr\noalign{\kern.5ex}
    \rightarrowfill\cr\noalign{\kern.5ex}
    \rightarrowfill\cr}}}\limits^{\!#1}}}

\newcommand{\factors}{\rightarrow}

\tikzset{%
    symbol/.style={%
        draw=none,
        every to/.append style={%
            edge node={node [sloped, allow upside down, auto=false]{$#1$}}}
    }
}

\newcommand{\tikzwrap}[2]{\tikzset{ampersand replacement=\&}\begin{tikzcd}#2#1\end{tikzcd}}

\NewDocumentCommand{\nattrans}{ m m O{""} O{""} O{""} }{%
      #1\arrow[rr, bend left=50, #3, ""{name=x, below}]
        \arrow[rr, bend right=50, #4, ""{name=y, above}]
\& \& #2\arrow[Rightarrow, from=x, to=y, #5]}

\NewDocumentCommand{\nattranstwo}{ m m m O{""} O{""} O{""} O{""} O{""} O{""} }{
      \nattrans{#1}{#2}[#4][#5][#8]
      \arrow[rr, bend left=50, #6 ""{name=xx, below}]
      \arrow[rr, bend right=50, #7, ""{name=yy, above}]
\& \& #3\arrow[Rightarrow, from=xx, to=yy, #9]}

\NewDocumentCommand{\natt}{ m m O{""} O{""} O{""} O{} }{
\tikzwrap{\nattrans{#1}{#2}[#3][#4][#5]}{#6}}

\NewDocumentCommand{\tonatt}{ m m m O{""} O{""} O{""} O{""} O{} }{
\tikzwrap{#1\arrow[r, #4] \& \nattrans{#2}{#3}[#5][#6][#7]}{#8}}

\NewDocumentCommand{\fromnatt}{ m m m O{""} O{""} O{""} O{""} O{} }{
\tikzwrap{\nattrans{#1}{#2}[#4][#5][#6]\arrow[r, #7] \& #3}{#8}}

\NewDocumentCommand{\nattwo}{ m m m O{""} O{""} O{""} O{""} O{""} O{""} }{
\tikzwrap{\nattranstwo{#1}{#2}{#3}[#4][#5][#6][#7][#8][#9]}{}}

\NewDocumentCommand{\nattwosmall}{ m m m O{""} O{""} O{""} O{""} O{""} O{""} }{
\tikzwrap{\nattranstwo{#1}{#2}{#3}[#4][#5][#6][#7][#8]}{[column sep=small]}}

\NewDocumentCommand{\natthree}{ m m O{""} O{""} O{""} O{""} O{""} O{} }{
\tikzwrap{
      #1\arrow[rrr, bend left=80, #3, ""{name=x, below}]
        \arrow[rrr, #4, ""{name=y, above}]
        \arrow[rrr, phantom, ""{name=yy, below}]
        \arrow[rrr, bend right=80, #5, ""{name=z, above}]
\& \& \& #2\arrow[Rightarrow, from=x, to=y, #6]
        \arrow[Rightarrow, from=yy, to=z, #7]
}{#8}}

\NewDocumentCommand{\rightleft}{ m m O{""} O{""} O{}}{
\tikzwrap{
    #1  \arrow[r, shift left=0.85ex, #3]
  \& #2 \arrow[l, shift left=0.85ex, #4]
}{#5}}

\NewDocumentCommand{\adjoint}{ m m O{""} O{""} O{} }{
\tikzwrap{
    #1  \arrow[r, shift left=0.62ex, phantom, ""{name=x,above}]
        \arrow[r, shift left=0.85ex, #3]
  \& #2 \arrow[l, shift left=0.42ex, phantom, ""{name=y,below}]
        \arrow[l, shift left=0.85ex, #4]
        \arrow[from=x, to=y, symbol={\scriptstyle\dashv}]
}{#5}}

\def\SURF{surf8}

\def\SAMPLEN{1008}
\def\SUBN{317}

\def\SAMPLET{61}

\def\SAMPLE{\SURF-sample\SAMPLEN-\SAMPLET}

\newcommand{\ARXIV}[1]{ #1 }
\newcommand{\SOCG}[1]{}

\title{A Theory of Sub-Barcodes}

\newcommand{\institute}{Department of Computer Science, North Carolina State University}

\author{Oliver A. Chubet}{\institute}{oliver.chubet@gmail.com}{https://orcid.org/0000-0002-4771-9894}{}
\author{Kirk P. Gardner}{\ }{k.gardner48@gmail.com}{https://orcid.org/0000-0001-5306-2174}{}
\author{Donald R. Sheehy}{\institute}{don.r.sheehy@gmail.com}{https://orcid.org/0000-0002-1825-0097}{}

\authorrunning{O. A. Chubet, K. P. Gardner, and D. R. Sheehy}
\Copyright{Oliver A. Chubet, Kirk P. Gardner, Donald R. Sheehy}



\funding{This work was partially funded by the NSF under grant CCF-2017980.}

\EventEditors{Oswin Aichholzer and Haitao Wang}
\EventNoEds{2}
\EventLongTitle{41st International Symposium on Computational Geometry
(SoCG 2025)}
\EventShortTitle{SoCG 2025}
\EventAcronym{SoCG}
\EventYear{2025}
\EventDate{June 23--27, 2025}
\EventLocation{Kanazawa, Japan}
\EventLogo{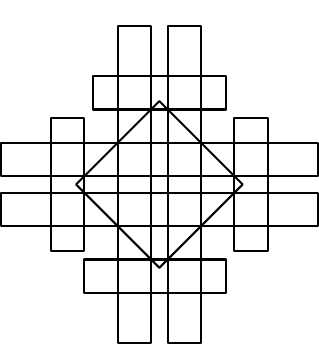}
\SeriesVolume{332}
\ArticleNo{XX}     

\nolinenumbers

\begin{document}

\maketitle


\begin{abstract}
  The primary tool in topological data analysis (TDA) is persistent homology, which involves computing a \emph{barcode}---often from point-cloud or scalar field data---that serves as a topological signature for the underlying function.
  In this work, we introduce \emph{sub-barcodes} and show how they arise naturally from \emph{factorizations} of persistence module homomorphisms.
  We show that, as a partial order induced by factorizations, the relation of being a sub-barcode is strictly stronger than the rank invariant,
  and we apply sub-barcode theory to the problem of inferring information about the barcode of an unknown Lipschitz function from samples.
  The advantage of this approach is that it permits strong guarantees---with no noise---while requiring no sampling assumptions, 
  and the resulting barcode is guaranteed to be a sub-barcode of every Lipschitz function that agrees with the data.
  We also present an algorithmic theory that allows for the efficient approximation of sub-barcodes using filtered Delaunay triangulations for Euclidean inputs.
\end{abstract}

\section{Introduction}\label{sec:introduction}

Topological data analysis (TDA) seeks to compute topological properties of unknown functions from finite samples.
The most widely-used tool for TDA is persistent homology (PH), and the most widely-studied functions are distance functions.
Thus, the algorithmic problems arising in this field tend to be either algebraic (PH computation), geometric (distance function representation), or both.

\begin{figure}[!ht]
    \centering
    \includegraphics[trim=200 150 200 150, clip, width=0.42\textwidth]{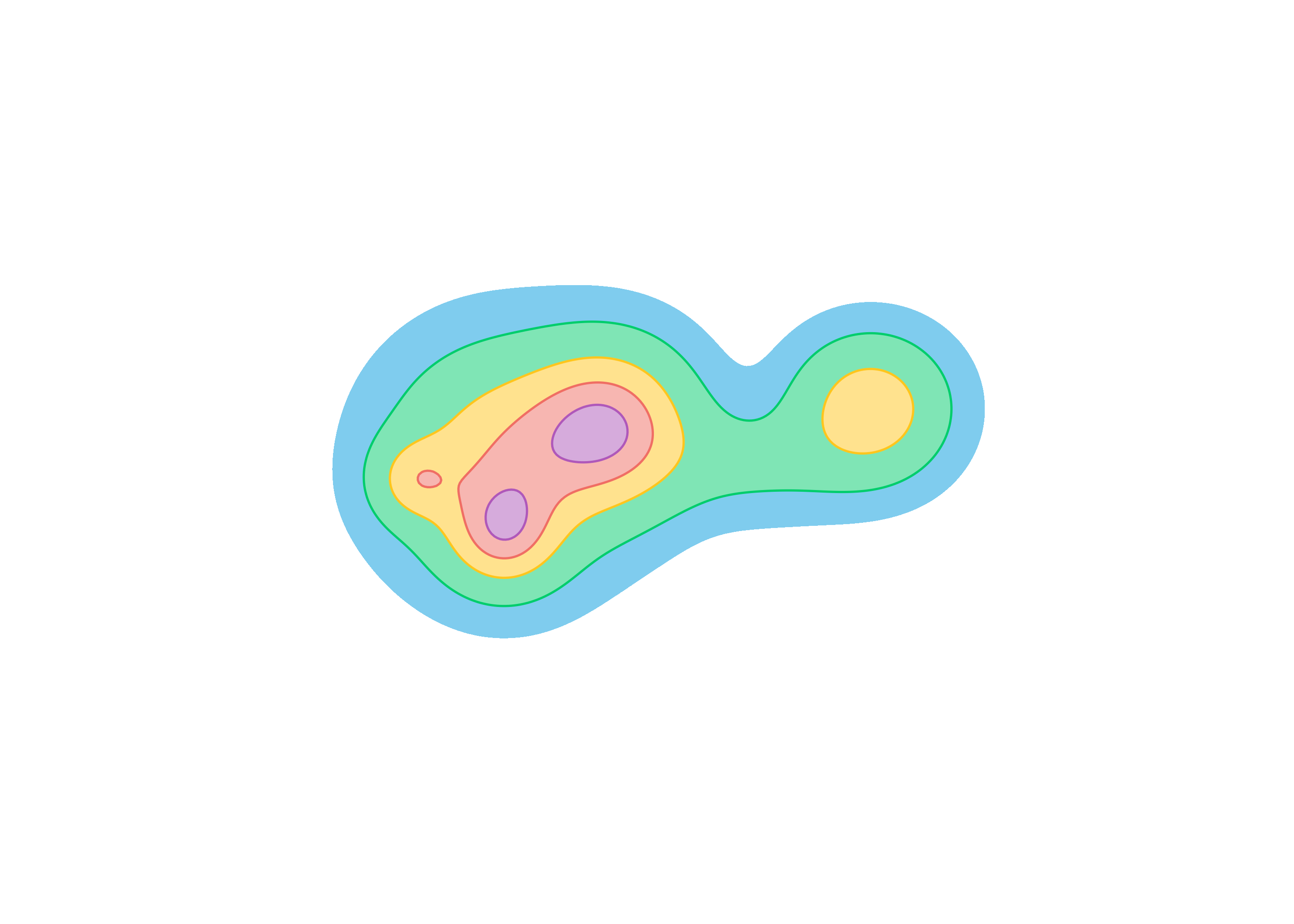}
    \includegraphics[width=0.55\textwidth]{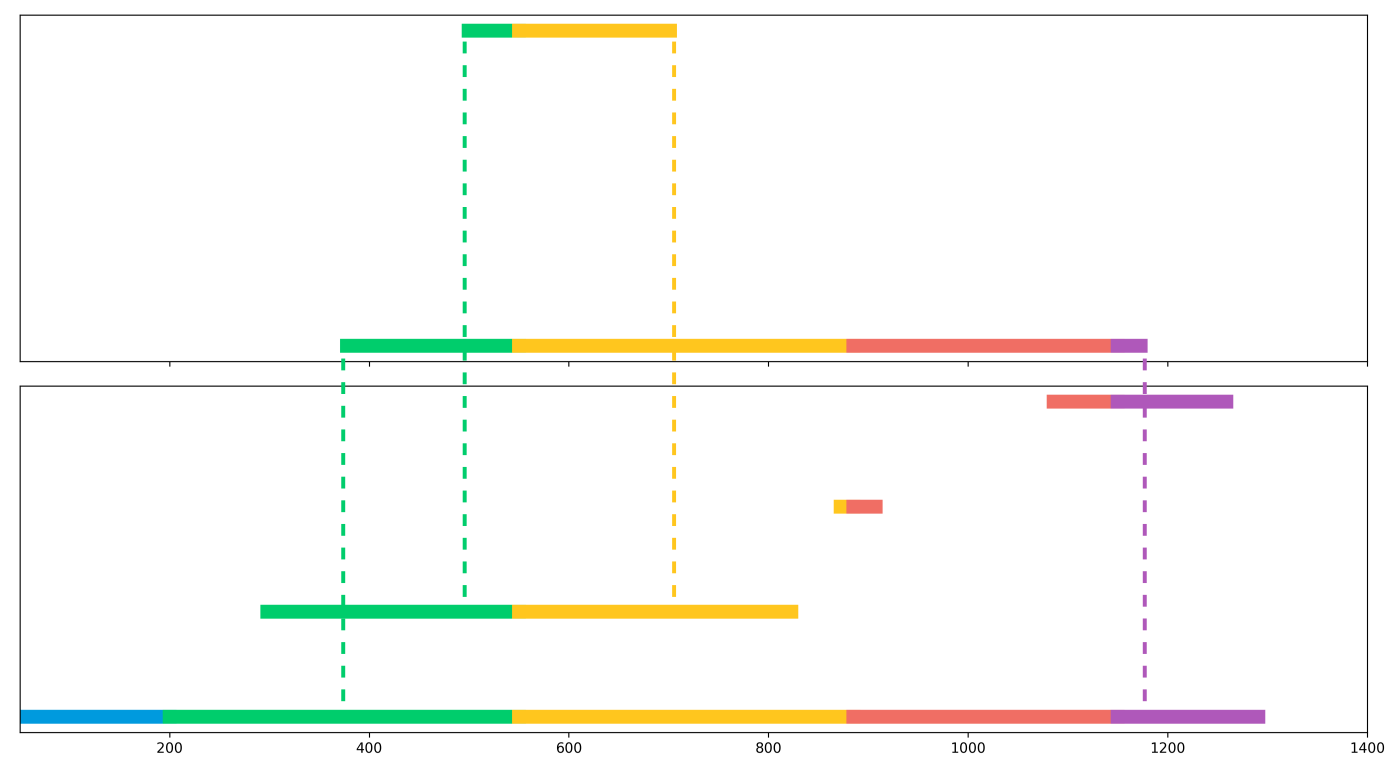}
    \caption{(Left) A contour plot of a Lipschitz function $f : X\to\RR$.
      (Right) The barcode $\bar(f)$ of $f$ (bottom) and a sub-barcode (top).
      All colors correspond to the function values indicated on the axis (bottom-right).}\label{fig:subbarcode}
\end{figure}

A PH pipeline starts with a real-valued function as input, and then considers the sublevel sets of this function.
The homology of the sublevels at each threshold value are tracked and aggregated into a \emph{barcode} (see Figure~\ref{fig:subbarcode}).
The goal is to compute or approximate this barcode, which often requires substantial assumptions on the quality of the sample.

In this paper, we relax this goal to instead compute whatever \emph{sub-barcode} is supported by the data we have.
We form a sub-barcode by taking a sub-interval of each bar in the barcode.
The formal definition of a sub-barcode and how it arises naturally in PH is covered in Section~\ref{sec:subbarcodes}.
The main idea is that, given \textit{some} of the data, we compute \textit{some} of the underlying structure.

In practice, analyzing the topology of a finite sample entails two main challenges:
\begin{itemize}
    \item \textbf{Extension:} If $S$ is a finite sample of the underlying space $X$, then we need to extrapolate the values for a function $f$ from the known values at $S$ to the unknown values at the rest of $X$.  We won't get the exact answer, but we would like some guarantees on the extension.
    \item \textbf{Discretization:} Having a good approximation to the function is insufficient: we also need a reliable discrete representation of the underlying space.
      In practice, these representations usually arise in the form of a simplicial complex or a simplicial set.
\end{itemize}

In the classic setting of distance functions in Euclidean space, strong sampling assumptions guarantee that the distance to a sample approximates the distance to the underlying set.
The Delaunay triangulation provides a perfect discretization, capturing the exact homology of the distance sublevels.
This was the original setting for PH~\cite{edelsbrunner02topological}.

Lipschitz functions generalize distance functions and allow us to represent a much wider class of inputs.
Several previous papers have shown that, given a sufficiently dense sample, a good approximation to the barcode of a Lipschitz function can be computed~\cite{buchet15topological,chazal11scalar,cohen2010lipschitz}.
The guarantees from such work is that the output barcode is close to the true barcode in a standard metric called the bottleneck distance.
Our approach eliminates all sampling assumptions, but gives guarantees in terms of a partial order on barcodes;
that is, it is guaranteed to produce no noise or spurious bars, but finds only those topological features (bars) that are sufficiently supported by the sample.

\paragraph*{Contributions and Outline}

The background and definitions appear in sections~\ref{sec:background} and~\ref{sec:subbarcodes}.
The \emph{algebraic theory} of sub-barcodes is presented in Section~\ref{sec:algebraic}.
First, we show how sub-barcodes are related to the widely-studied rank invariant.
Surprisingly, despite being a complete invariant for persistence, the natural extension of the rank invariant to a partial order is strictly less discriminating than sub-barcodes.
This would be useless if it were not the case that sub-barcodes can be computed, a fact we show in the main sub-barcode theorem (Theorem~\ref{thm:subbarcodes}),
which states that sub-barcodes arise whenever we have one persistence module homomorphism factoring through another.

\begin{figure}[!ht]
    \centering
    \includegraphics[width=\textwidth]{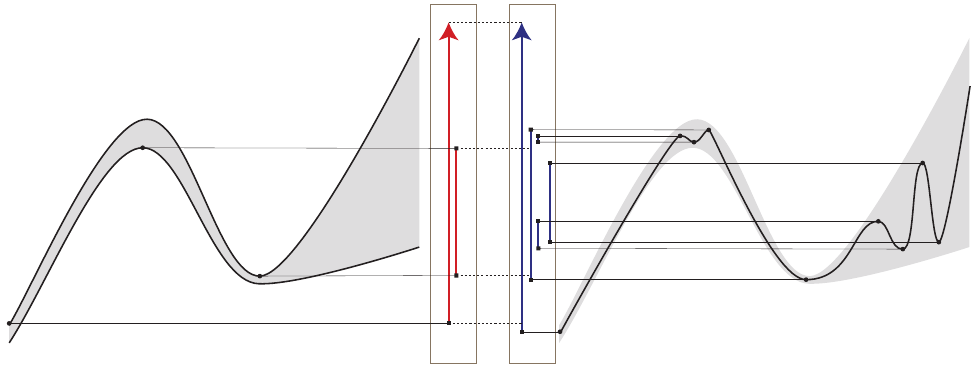}
    \caption{On the left, two functions are depicted, one is an upper bound and the other is a lower bound on an unknown function $f:\R \to \R$.
      There is a corresponding barcode associated with the pair that matches minima in the upper bound to maxima in the lower bound.
      On the right is a candidate function $f$ that lies between the upper and lower bounds and its barcode $\bar(f)$.
      The barcode of the inclusion of the upper and lower bounds is a sub-barcode of $\bar(f)$.\label{fig:upper_and_lower_bounds}
      }
\end{figure}

In Section~\ref{sec:extension}, we lay out the \emph{extension theory}.
According to the algebraic theory of the previous section, the key to producing sub-barcodes is to find suitable upper and lower bounds on the unknown function.
We show that maximum and minimum Lipschitz extensions serve this purpose (Figure~\ref{fig:upper_and_lower_bounds}),
as well as how to give good guarantees with only an approximation to the upper and lower bounds (Corollary~\ref{cor:lipschitz_subbarcode}).
This section also contains a much more general result to encompass other classes of functions beyond Lipschitz functions (Theorem~\ref{thm:beyond}).

The \emph{discretization theory} is presented in Section~\ref{sec:discretization}.
Here, we show how to approximate the Lipschitz extension sub-barcode of the previous section using Delaunay triangulations.
This leads to a theory of semi-supervised TDA in which only a subset of function values are known (Theorem~\ref{thm:semi_supervised}).
We also give a tighter approximation using the barycentric decomposition of the Delaunay triangulation (Theorem~\ref{thm:bary}),
and show that this method also applies to a wider class of functions (Example~\ref{ex:alpha}).

The \emph{categorical theory} of sub-barcodes is presented in Section~\ref{sec:categorical}.
The TDA and PH pipeline depend on category theory to express the way relationships between inputs impact relationships between outputs, i.e., barcodes.
We relate sub-barcodes to the classic approach of smoothing barcodes and show how this arises naturally from categorical considerations.
We also give a new perspective on barcodes in general for which sub-barcodes are shown to be true subobjects in the categorical sense,
and discuss how this construction is related to the theory of fuzzy sets~\cite{goguen67fuzzy,barr86fuzzy,spivak09metric}.

\section{Background}\label{sec:background}

\paragraph*{Functions and Filtrations}

Let $(X,\dist)$ be a metric space.
When the metric is Euclidean, we denote the distance as a norm, i.e., $\dist(x,y):=\|x-y\|$.
The distance from a point $x$ to a (finite) set $U$ is denoted $\dist(x,U):= \min_{u\in U}\dist(x,u)$.

The real-valued functions $f:X\to\R$ form a vector space where $(f+\alpha\,g)(x):= f(x) + \alpha\,g(x)$, for all $\alpha \in \R$.
We also have a poset of functions, where $f\ge g$ iff $f(x)\ge g(x)$ for all $x\in X$.
A function $f:X\to \R$ is \emph{$\lambda$-Lipschitz} if, for all $x,y\in X$, we have 
\[
    f(y) - \lambda\;\dist(x,y) \le f(x) \le f(y) + \lambda\;\dist(x,y).
\]
If $\lambda$ is not specified, then it is assumed that $\lambda=1$.
This is perhaps not the most common way to express the Lipschitz condition, but it is closest to how it is used in this work.

The \emph{$t$-sublevel set} $\Sub_f(t)$ of $f$ is the set of points $x\in X$ with $f(x)\leq t$:
\[
    \Sub_f(t) := f^{-1}((-\infty,t]) = \{x\in X\mid f(x) \le t\}.
\]
This notation is designed to indicate that $\Sub_f$ is a \emph{filtration}: for $s\le t$ we have $\Sub_f(s)\subseteq \Sub_f(t)$.
There is a partial order on sublevel filtrations where $\Sub_f\subseteq \Sub_g$ iff $\Sub_f(t)\subseteq \Sub_g(t)$ for all $t\in \R$.
It is a simple exercise to check that, if $f\ge g$, then $\Sub_f \subseteq \Sub_g$; i.e., larger functions have smaller sublevel sets.

\begin{remark}
  The mapping from a function to its sublevel filtration is a contravariant functor from the poset of functions to the poset of filtrations.
  The filtrations themselves are functors $\RR\to \Top$ from the totally ordered set $\RR$ of real numbers to $\Top$, the category of topological spaces.
  Throughout, we use remarks like this one to highlight the categorical interpretations.
  The reader who is unfamiliar with these terms may still follow all definitions and algorithms.
\end{remark}

\paragraph*{Persistence Modules}

A \emph{pointwise finite dimensional (p.f.d.) persistence module} $M:\R\to \vec$ assigns a (finite-dimensional) vector space $M(t)$ to all $t\in \RR$ and a linear map $M_{s\le t}: M(s)\to M(t)$ for all $s\le t$.
The map $M_{s\le s}$ is the identity on $M(s)$, and for all $s\le t\le u$, $M_{s\le u} = M_{t\le u}\circ M_{s\le t}$.
For persistence modules $M$ and $N$, a homomorphism $\phi:M\to N$ assigns linear maps $\phi_t:M(t)\to N(t)$ for all $t\in \R$ so that for all $s\le t$, we have $\phi_t\circ M_{s\le t} = N_{s\le t}\circ \phi_s$.

\begin{remark}
  A persistence module $M$ is a functor $M : \RR\to \vec$, and a persistence module homomorphism $M\to N$ is a natural transformation $\phi : M\Rightarrow N$ between functors $\RR\to \vec$,
  so that the category of persistence modules is precisely the category $\vec^\RR = [\RR,\vec]$ of functors $\RR\to \vec$.
\end{remark}



A \emph{factorization} of homomorphisms $F : M\to N$ to $G:M'\to N'$ is given by a pair of homomorphisms $\varphi = \big(\varphi_1 : M'\to M,\ \varphi_2: N\to N'\big)$ such that $G = \varphi_2\circ F\circ \varphi_1$ (Diagram~\ref{dgm:mod_fact}).
\begin{equation}\label{dgm:mod_fact}
  \begin{tikzcd}[sep=large]
      M \arrow[d,"F"]
    & M'\arrow[l,"\varphi_1"']
        \arrow[d,"G"]
    \\
      N \arrow[r,"\varphi_2"]
    & N'.
  \end{tikzcd}
\end{equation}

Let \Fact{\vec[\RR]} denote the category with homomorphisms of persistence modules as objects, and arrows $\varphi : F\factors G$.
Then $\varphi: F\factors G$ is a factorization of $G$ through $F$.



\begin{example}
  Letting $M(\epsilon)$ denote a persistence module $M$ shifted by a constant $\epsilon\in\RR$, a $2\epsilon$-interleaving is traditionally given by a pair of commuting diagrams


\vspace{3ex}
\begin{minipage}{0.47\linewidth}
  \begin{equation}\label{dgm:left_fact}
    \begin{tikzcd}
      M(-\epsilon)
        \arrow[rr,"M\e"]
        \arrow[dr,"\Phi(-\epsilon)"'] & &
      M(\epsilon)\\
      & N
        \arrow[ur,"\Psi"'] &
    \end{tikzcd}
  \end{equation}
\end{minipage}
\begin{minipage}{0.47\linewidth}
  \begin{equation}\label{dgm:right_fact}
    \begin{tikzcd}
      & M
        \arrow[dr,"\Phi"] & \\
      N(-\epsilon)
        \arrow[rr,"N\e"']
        \arrow[ur,"\Psi(-\epsilon)"] & &
      N(\epsilon)
    \end{tikzcd}
  \end{equation}
\end{minipage}
\vspace{2ex}

  \noindent depicting factorizations of homomorphisms $M\e : M(-\epsilon)\to M(\epsilon)$ and $N\e : N(-\epsilon)\to N(\epsilon)$ by a pair
  $\big( \Phi : M\to N(\epsilon),\ \Psi: N\to M(\epsilon) \big).$
  In this section, we show that the commutativity of Diagrams~\eqref{dgm:left_fact} and~\eqref{dgm:right_fact} imply sub-barcode relations $\bar(M\e)\subb \bar(N)$ and $\bar(M)\sqsupseteq\bar(N\e)$.
\end{example}

\paragraph*{Homology}

\emph{Homology} is a topological invariant that is particularily useful because it is both informative and efficiently computable for triangulable topological spaces.
A homology functor $\H : \Top\to \vec$ associates a vector space $\H(X)$ to a topological space $X$, and takes continuous maps between topological spaces to linear maps between the corresponding vector spaces.

Persistence modules naturally arise in TDA by taking the homology of a sublevel filtration of a function.
This assembles the functorial TDA pipeline from functions to filtrations to persistence modules.
It is useful to see these as functors and not just functions because the additional structure of functoriality tells us that, 
if $f\ge g$, then there is a persistence module homomorphism $\H\;\Sub_f\to \H\;\Sub_g$.
As we will see, these homomorphisms can be combined into factorizations (leading to sub-barcodes) and interleavings (leading to approximations).

\begin{figure}[!ht]
  \centering
  \includegraphics[trim=200 150 200 150, clip, width=0.32\textwidth]{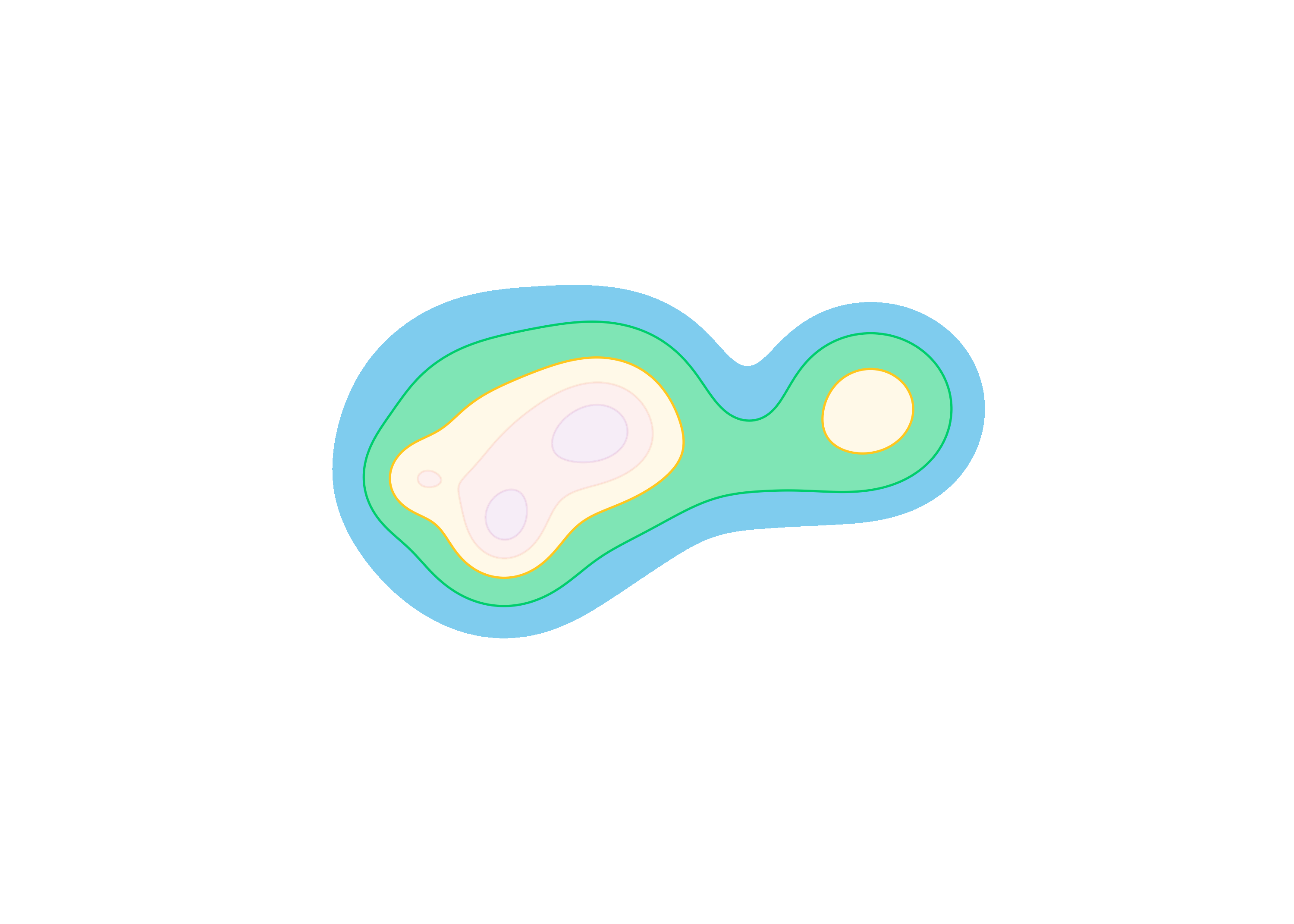}
  \includegraphics[trim=200 150 200 150, clip, width=0.32\textwidth]{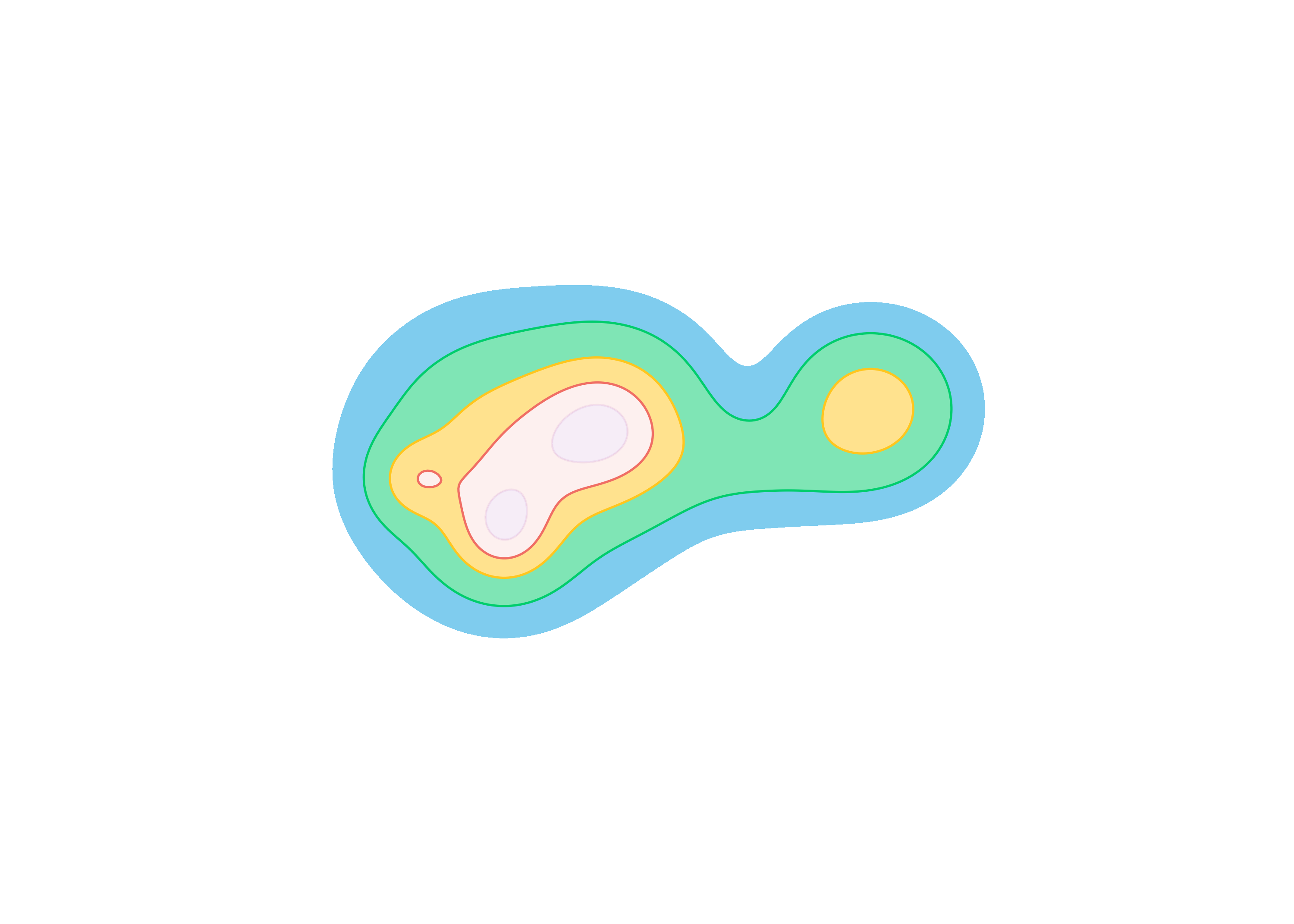}
  \includegraphics[trim=200 150 200 150, clip, width=0.32\textwidth]{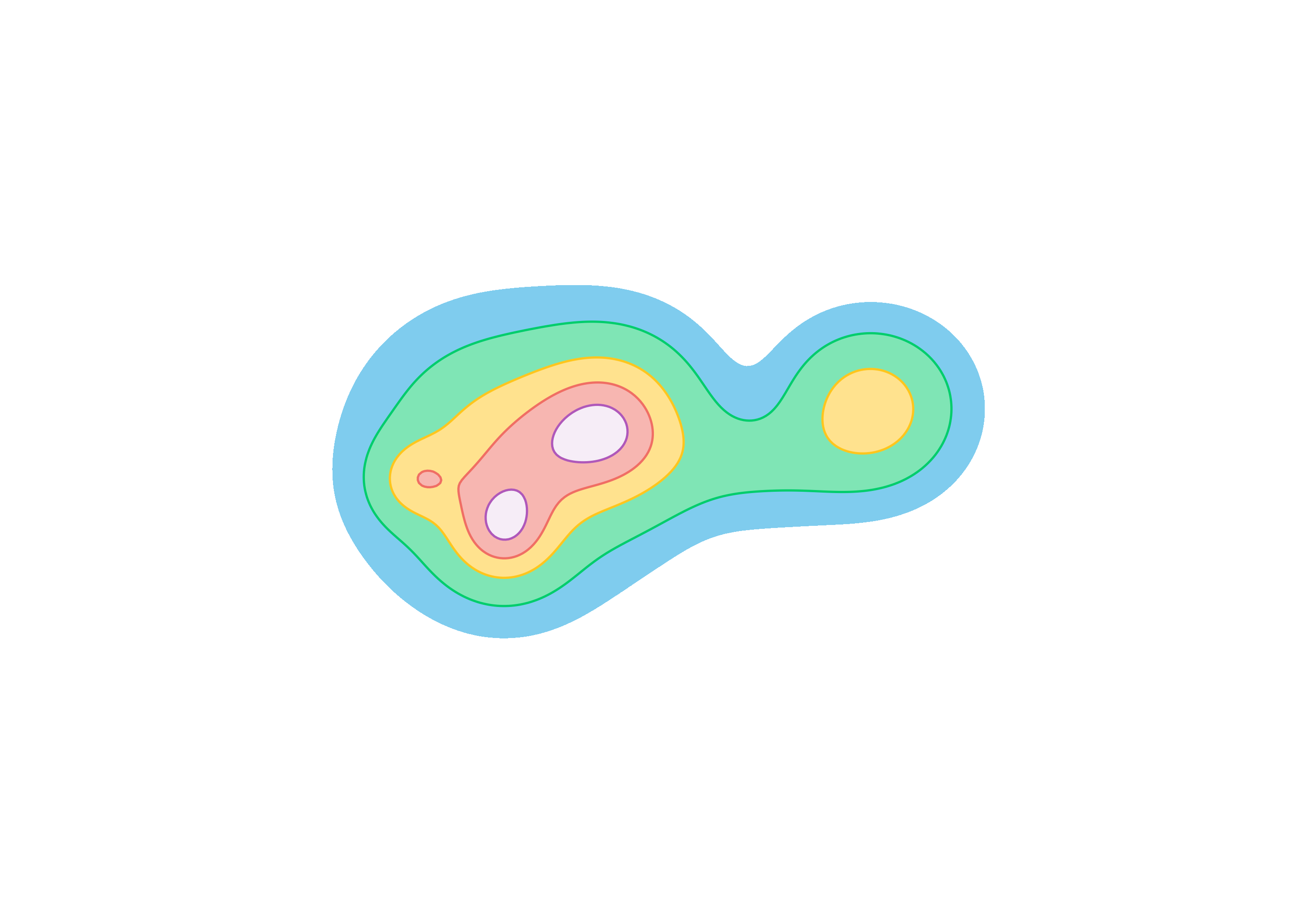}
  \includegraphics[width=\textwidth]{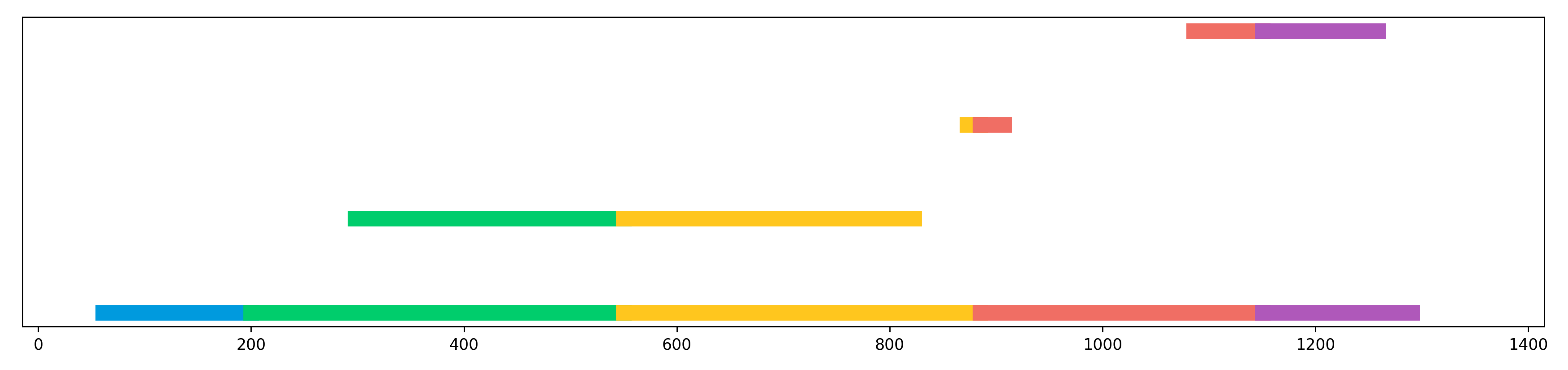}
  \caption{Sublevels of a 3-Lipschitz function $f : X\to\RR$ on a subset of $\RR^2$ (top) and its barcode $\bar(f) = \bar(\H_1\Sub_f)$ in dimension 1 (bottom).
Colors correspond to the function values indicated on the bottom-axis.}\label{fig:scalar_field_barcode}
\end{figure}

\section{Barcodes and Sub-Barcodes}\label{sec:subbarcodes}

Let $\I$ denote the poset of intervals on $\R$ ordered by inclusion.
A \emph{barcode} $B$ is a function $B:\ul{B}\to \I$ where $\ul{B}$ is a set whose elements $\beta\in\ul{B}$ are called \emph{bars}.
The barcode assigns an interval to each bar.
Note that $B$ is not necessarily injective.
When there is no confusion, we identify the bars in a barcode with their corresponding intervals.
A \emph{sub-barcode mapping} between barcodes $A$ and $B$ is a function $\phi:\ul{A}\to \ul{B}$ such that, for all $\alpha\in \ul{A}$, we have $A(\alpha)\subseteq B(\phi(\alpha))$.
If $\phi$ is injective, then we call it a \emph{sub-barcode matching} and say $A$ is a \emph{sub-barcode} of $B$, denoted $A\subb B$.
That is, $A$ is a sub-barcode of $B$ if $A$ can be formed from $B$ by taking a subinterval of each bar in a subset of $\ul{B}$ (Figure~\ref{fig:many_hats}).

There is a barcode $\bar(M)$ corresponding to any persistence module $M$.
Note, the image of a persistence module homomorphism $\phi:M\to N$ is also persistence module.
We overload notation and write $\bar(\phi)$ to indicate $\bar(\im\phi)$.


\begin{figure}[ht]
\input{tikz/colors}
\begin{center}
\scalebox{2}{
\begin{tikzpicture}[thick]
   \filldraw[gray!15] (-3.5,1.2) rectangle (2.25,0.4);
   \draw (-3.5,0.8) node[black,left] {\tiny$A$};
     \draw[\ca] (-1,1) -- (1,1); 
     \draw[\cb] (-1.5,0.8) -- (1.5,0.8); 
     \draw[\cc] (-2.5,0.6) -- (-0.5,0.6); 
   \filldraw[gray!15] (-3.5,0.2) rectangle (2.25,-0.8);
   \draw (-3.5,-0.3) node[black,left] {\tiny$B$};
   \draw[\ca] (-1.75,0) -- (1.25,0); 
   \draw[\cb] (-2,-0.2) -- (1.75,-0.2); 
   \draw[\cc] (-2.75,-0.4) -- (0,-0.4); 
   \draw[opacity=0.66] (-3,-0.6) -- (-2.1,-0.6); 

   \draw[\ca,thin,dashed] (-1,1) -- (-1,0) (1,1) -- (1,0); 
   \draw[\cb,thin,dashed] (-1.5,0.8) -- (-1.5,-0.2) (1.5,0.8) -- (1.5,-0.2); 
   \draw[\cc,thin,dashed] (-2.5,0.6) -- (-2.5,-0.4) (-0.5,0.6) -- (-0.5,-0.4); 

 \end{tikzpicture}
}
\end{center}
    \caption{Two barcodes $A$ and $B$ and a sub-barcode matching $M : A\to B$.}\label{fig:many_hats}
\end{figure}

From the PH pipeline, we have persistence modules $\H\;\Sub_f$ associated with each continuous function $f : X\to \RR$.
We again simplify our notation and write $\bar(f)$ to denote $\bar(\H\;\Sub_f)$ and $\bar(f\ge g)$ to denote $\bar(\im \H(\Sub_f\subseteq \Sub_g))$.
The object giving us a barcode is always clear from context.


We now have enough definitions to state the main theorem of sub-barcodes.
The proof is postponed until Section~\ref{sec:algebraic} when we will have developed some necessary tools.

\begin{theorem}[The Sub-Barcode Theorem]\label{thm:subbarcodes}
    If there exists a factorization $\varphi:F\factors G$ of persistence module homomorphisms (i.e., $G = \varphi_2 F \varphi_1$), then $\bar(G)\subb \bar(F)$.
\end{theorem}

\section{The Algebraic Theory}\label{sec:algebraic}

\subsection{Sub-barcodes vs. Ranks}

Let $\rk~m$ denote the \emph{rank} of a linear map $m$ defined as the dimension of its image, $\im m$.
The \emph{rank invariant} $\rank_M$ of a persistence module $M : \RR\to\vec$ is a function from ordered pairs in $\RR$ to the natural numbers defined as $\rank_M(s\le t):= \rk M_{s\le t}$.
It is straightforward to extend this invariant to homomorphisms $F:M\to N$ by letting $\rank_{F}(s\le t):= \rk(F_t \circ M_{s\le t})$.
In the following, we consider the rank invariant for a persistence module $M$ to be the rank invariant of its identity homomorphism, i.e., $\rank_M := \rank_{\id_M}$.

A basic fact from linear algebra is that, if a linear map $g$ factors through a second linear map $f$, then $\rk(g) \le \rk(f)$.
It follows immediately that, if there is a factorization $F\factors G$ of persistence module homomorphisms, then for all $s\leq t$,
\[
    \rank_G(s\le t) \le \rank_F(s\le t).
\]
This puts a partial order on rank invariants that we call the \emph{sub-rank relation} and denote by $\rank_G \le \rank_F$.

\begin{remark}
    $\rk$ is a contravariant functor $\Fact{\vec}\to \NN$ and $\rank$ is a contravariant functor $\Fact{\vec^\RR}\to \NN$, i.e., from factorizations of vector spaces (resp. persistence modules) to the poset of natural numbers.
    An ordering of two rank invariants is a natural transformation between these functors.
\end{remark}

In light of the sub-barcode theorem, the rank invariant is the same \emph{type} of invariant as sub-barcodes.
It puts a partial order on morphisms induced by factorizations.
However, although the barcode can be constructed from the rank invariant, the natural ordering of rank invariants is a weaker invariant than sub-barcodes as expressed in the following.

\begin{theorem}\label{thm:subbarcodes_vs_ranks}
    The sub-barcode order is a more discriminating invariant than the sub-rank order in the sense that,
    for all persistence module homomorphisms $F$ and $G$, $\bar(G) \subb \bar(F)$ implies that $\rank_{G}\le \rank_{F}$, but not vice-versa.
\end{theorem}
\begin{proof}
    The rank invariant is easily extracted from a barcode: $\rank_{F}(s\le t)$ is the number of bars in $\bar(F)$ that contain the closed interval $s\le t$.
    It follows immediately from the injectivity of the sub-barcode matching (and the pigeonhole principle) that $\bar(G)\subb \bar(F)$ implies that $\rank_{G}\le \rank_{F}$.

    For the other side of the proof, it suffices to give an example of persistence modules $M$ and $N$ such that $\rank_M\le \rank_N$ while $\bar(M)\not\subb \bar(N)$.\footnote{As before, $\rank_M$ is shorthand for $\rank_{\id_M}$.}
    Let $M$ be a persistence module where $\bar(M)$ has two bars corresponding to intervals $[0,1)$ and $[2,3)$,
    and let $N$ be a persistence module such that $\bar(N)$ has only one bar corresponding to the interval $[0,3)$.
    Clearly, $\rank_M\le \rank_N$, but it is not possible for a barcode with two bars to be a sub-barcode of one with only one bar.
    Thus $\bar(M)\not\subb \bar(N)$.
\end{proof}

\subsection{Induced Matching Theory Proof of the Sub-Barcode Theorem}

Although not expressed in these terms, Bauer and Lesnick's theory of induced matchings~\cite{bauer13induced} shows that sub-barcode matchings arise from certain homomorphisms.
The induced matchings in that work are matchings between the bars that do not (necessarily) satisfy the inclusion requirement of a sub-barcode matching.
They showed that for any homomorphism $F:M\to N$, there is a partial bijective function (a \emph{matching}) between the bars with the property that every bar of $\bar(F)$ is the intersection of the intervals of the matched bars from $\bar(M)$ to $\bar(N)$.
These are constructed from the pair of canonical (functorial) matchings $\bar(M)\to \bar(F)\to \bar(N)$ induced by the epi-mono factorization of $F$ through its image.
Translated into the vocabulary of sub-barcodes, this result implies that $\bar(F) \subb \bar(M)$ and $\bar(F) \subb \bar(N)$.

More generally, if $F$ is a monomorphism (all maps $F_t$ are injective), then $\bar(M)\subb \bar(N)$.
This is because the induced matching of a monomorphism matches every bar, in addition to the property that, if $[a,b]$ is matched to $[c,d]$, then $a\ge c$ and $b=d$;
thus, a submodule relation between persistence modules implies a sub-barcode relation between the barcodes.

If $F$ is an epimorphism (all maps $F_t$ are surjective), then the induced matching is surjective, and if $[a,b]$ is matched to $[c,d]$, then $a=c$ and $d\le b$.
So, reversing this induced matching, we get $\bar(N)\subb \bar(M)$.
We use these matchings in the proof of the Sub-Barcode Theorem below.


\begin{proof}[Proof of The Sub-Barcode Theorem (Theorem~\ref{thm:subbarcodes})]
    The factorization $\varphi = (\varphi_1,\varphi_2):F \factors G$ entails $G = \varphi_2F\varphi_1$.
    Let $m:\im G\into \im \varphi_2F$ be the unique monomorphism given by the universal property of images,
    and let $e:\im F \twoheadrightarrow \im \varphi_2F$ be the epimorphism given by resticting $\varphi_2$ to $\im F$.
    Using the induced matching theory, $m$ gives a sub-barcode matching $\bar(\varphi_2F)\subb \bar(F)$.
    Then, the reverse of the matching induced by $e$ gives a sub-barcode matching $\bar(G)\subb \bar(\varphi_2F)$.
    Putting these together, we get
    \[
        \bar(G)\subb \bar(\varphi_2F) \subb \bar(F).\qedhere
    \]
\end{proof}


The following corollary for ordered sequences of functions forms the basis for all of our applications of sub-barcodes in TDA.

\begin{corollary}\label{cor:subbarcodes}
    If $h\ge g\ge f\ge e$ are functions $X\to \RR$, then $\bar(h\ge e)\subb \bar(g\ge f)$.    
\end{corollary}
\begin{proof}
    It suffices to observe that the ordering on the functions gives a factorization $(g\ge f) \to (h\ge e)$ of $h\ge e$ through $g\ge f$.
    By the functoriality of the sublevel functor and homology, we get the corresponding factorization of persistence modules.
    The claim then follows from  Theorem~\ref{thm:subbarcodes}.
\end{proof}

\section{The Extension Theory}\label{sec:extension}

Given an unknown Lipschitz function $f:X\to \R$ and a sample $S\subset X$, let $f_S$ denote the function $f$ restricted to the points of $S$.
The pair $(S,f_S)$ denotes input to a TDA problem where the goal is to infer as much as possible about $f$.
Any function $g:X\to \RR$ that agrees with $f_S$ at the points of $S$ is called an \emph{extension} of $f_S$.
If $g$ is also Lipschitz, then it is called a \emph{Lipschitz extension}.
By the definition of Lipschitz continuity, each point $s\in S$ puts an upper bound and a lower bound on the unknown $f$ so that
\[
    f(s) - \dist(x,s) \le f(x) \le f(s) + \dist(x,s).
\]
The combination of these upper and lower bounds gives the \emph{maximum and minimum Lipschitz extensions}, defined respectively for each $x\in X$ as
\[
    \check{f}_S(x) := \min_{s\in S}f(s) + \dist(x,s) \text{~~and~~} \hat{f}_S(x) := \max_{s\in S}f(s) - \dist(x,s).
\]
Note that the max extension is the minimum of the upper bounds and vice versa (see also~\cite{lawvere73metric}).

The following theorem states that, for every Lipschitz function $f$ that agrees with the data, the max and min Lipschitz extensions yield a barcode that is contained in $\bar(f)$.

\begin{theorem}\label{thm:lipschitz_subbarcode}
    Given $S\subseteq X$ and $f_S:S\to \RR$, for all Lipschitz extensions $f$ of $f_S$, we have $\bar(\check{f}_S \ge \hat{f}_S) \subb \bar(f)$.
\end{theorem}
\begin{proof}
    It suffices to observe that for all Lipschitz extensions of $f$, we have $\check{f}_S \ge f \ge \hat{f}_S$.
    Because $\bar(f) = \bar(f\ge f)$, the result follows from Corollary~\ref{cor:subbarcodes}.
\end{proof}

\paragraph*{The Choice of Lipschitz Constants.}

It is natural to ask whether (and how) the choice of the Lipschitz constant affects this result.
The previous statements have all assumed $1$-Lipschitz functions.
More generally, tuning the Lipschitz constant naturally leads to a filtration of the barcode itself as shown in the following.

\begin{corollary}\label{cor:lipschitz_subbarcode}
    For any $t\in \R$, let $\check{f}_S^{(t)}$ and $\hat{f}_S^{(t)}$ denote the (resp.) max and min $t$-Lipschitz extensions.
    If $f_S$ is $t_0$-Lipschitz and $t_0\le t_1\le t_2$, then $\bar(\check{f}_S^{(t_1)}\ge \hat{f}_S^{(t_1)}) \subb \bar(\check{f}_S^{(t_2)}\ge \hat{f}_S^{(t_2)})$.
\end{corollary}

\paragraph*{A Sub-Barcode that is Close in Bottleneck Distance.}

A fundamental challenge in computing sub-barcodes from Lipschitz extensions is that it is not clear how to construct a simplicial complex that can represent the sublevel sets of both the max and min extension at the same time.
So, in order to still derive guarantees, it will suffice to have approximations to these extensions.
The following theorem shows that a bottleneck-close sub-barcode is possible if we have close approximations to the upper and lower bounds.

\begin{theorem}
    Let $u' \ge u \ge f \ge \ell \ge \ell'$ be an ordered sequence of functions $X\to \RR$ such that $\|\ell - \ell'\|_{\infty}\le \e$ and $\|u - u'\|\le \e$.
    Then, $\dist_B\big(\bar(u\ge \ell), \bar(u'\ge \ell')\big)\le \e$.
\end{theorem}
\begin{proof}
    The proof is a straightforward application of the general stability theorem for image persistence.
    The bounds $\|\ell - \ell'\|_{\infty}\le \e$ and $\|u - u'\|\le \e$ imply that $\ell' + \e \ge \ell \ge \ell'$ and $u'\ge u \ge u'-\e$.
    As all maps from ordered functions induce inclusions of sublevel sets, all required maps commute to form an interleaving in the arrow category of filtrations.
    This gives an interleaving in the arrow category of persistence modules which leads to an $\e$-interleaving of the image persistence modules.
\end{proof}

\paragraph*{Beyond Lipschitz Functions.}

The careful reader may note that there is nothing particularly special about Lipschitz functions in the treatment above.
One could consider other classes of functions as well.
More generally, it suffices to have an upper and lower bound on the unknown function.
In the Lipschitz case, the Lipschitz extensions form a lattice of hypotheses that we are considering.
Recall that a poset is a lattice if every subset has a join (least upper bound) and a meet (greatest lower bound).
The symbol $\join$ is used for joins and $\meet$ is used for meets.
This is the origin of our notation $\check{f}_S$ and $\hat{f}_S$.

For a set $\F$ of functions, let $\Join \F:= \Join_{f\in \F}f$ and $\Meet\F:= \Meet_{f\in \F}f$.
In particular, if $\F$ denotes the set of Lipschitz extensions of $f_S$, then $\check{f}_S = \Join \F$ and $\hat{f}_S = \Meet\F$.
For other lattices of hypotheses, the same proof as for the Lipschitz case (Theorem~\ref{thm:lipschitz_subbarcode}) implies the following theorem.

\begin{theorem}\label{thm:beyond}
    Given a lattice $\F$ of functions $X\to \RR$, for all $f\in \F$, we have 
    \[ \bar\big(\Join \F\ge \Meet \F\big)\subb \bar(f).\]
\end{theorem}

\section{The Discretization Theory}\label{sec:discretization}

The preceding sections laid out the theory of sub-barcodes for persistence modules, as well a the useful setting of Lipschitz functions.
The following section shows how to approximate these barcodes.

The fundamental challenge is to provide a single discretization of space that approximates (or at least bounds) both Lipschitz extensions at the same time.
We assume throughout this section that $X$ is a convex polytope in $\RR^d$.\footnote{The convexity of $X$ can be relaxed in many settings. Assuming $X$ is convex simplifies our use of the Nerve theorem.}
For a finite set $S\subset X$, the \emph{Voronoi diagram} assigns a convex polytope called the \emph{Voronoi cell} to each point $s\in S$ defined as
\[
    \vor_S(s):= \{x\in X\mid \dist(x,s) = \dist(x,S)\}.
\]
The \emph{radius} of $\vor_S(s)$ is the maximum distance from $s$ to any point in the cell: $r_s:= \max_{x\in \vor_S(s)}\dist(x,s)$.
More generally, for $\sigma \subseteq S$, we have a Voronoi cell
\[
    \vor_S(\sigma):= \bigcap_{s\in \sigma}\vor_S(s).
\]

Dually, the \emph{Delaunay triangulation} $\del_S$ associated with $S$ is the simplicial complex with vertex set $S$ and simplices $\del_S = \{\sigma\subseteq S\mid \vor_S(\sigma)\neq \emptyset\}$.
The Delaunay triangulation is widely used for persistent homology on Euclidean inputs.


\subsection{Simple Voronoi Extensions and a Delaunay Filtration}

\begin{figure}
    \includegraphics[width=0.45\textwidth]{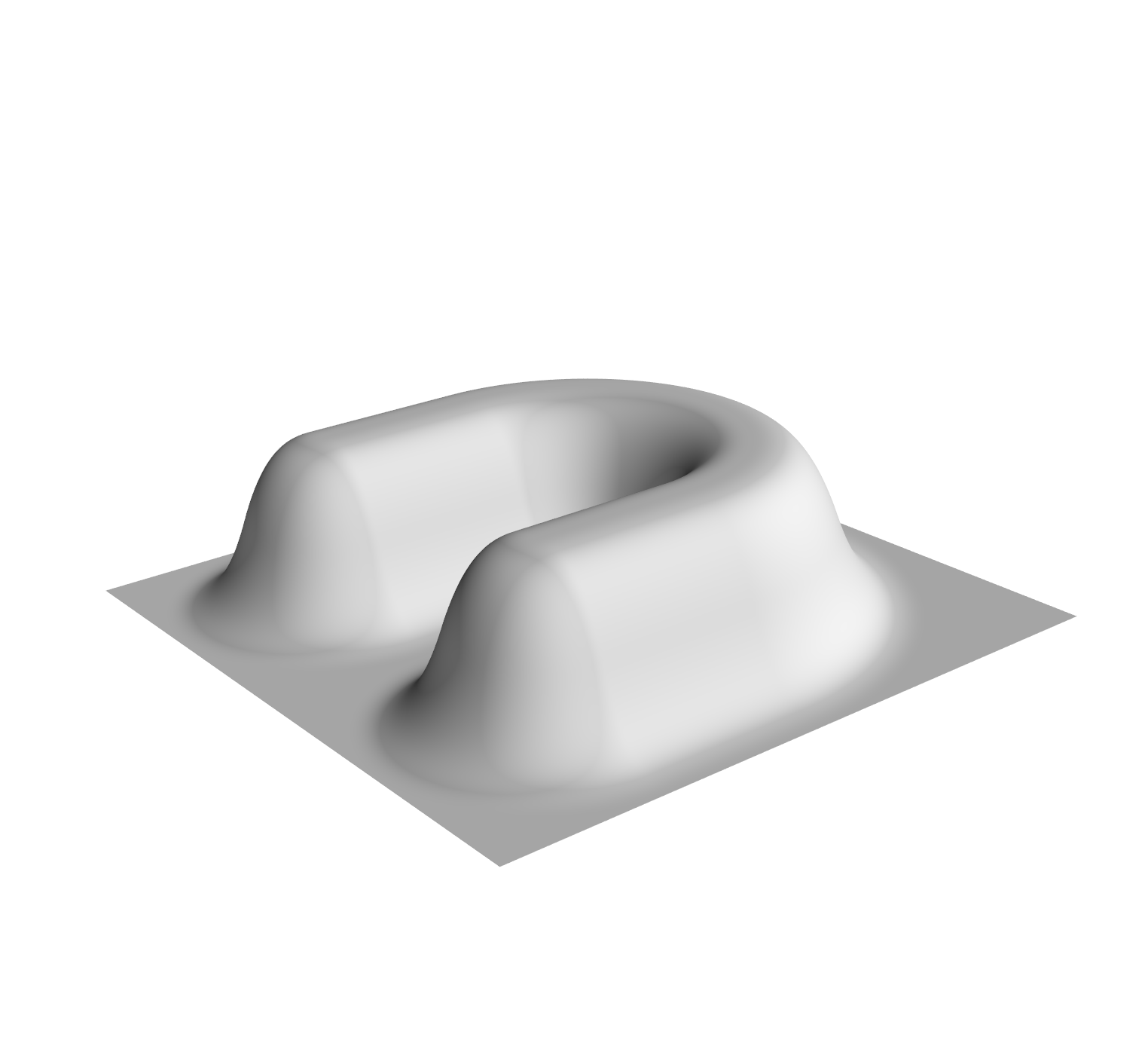}
    \includegraphics[width=0.45\textwidth]{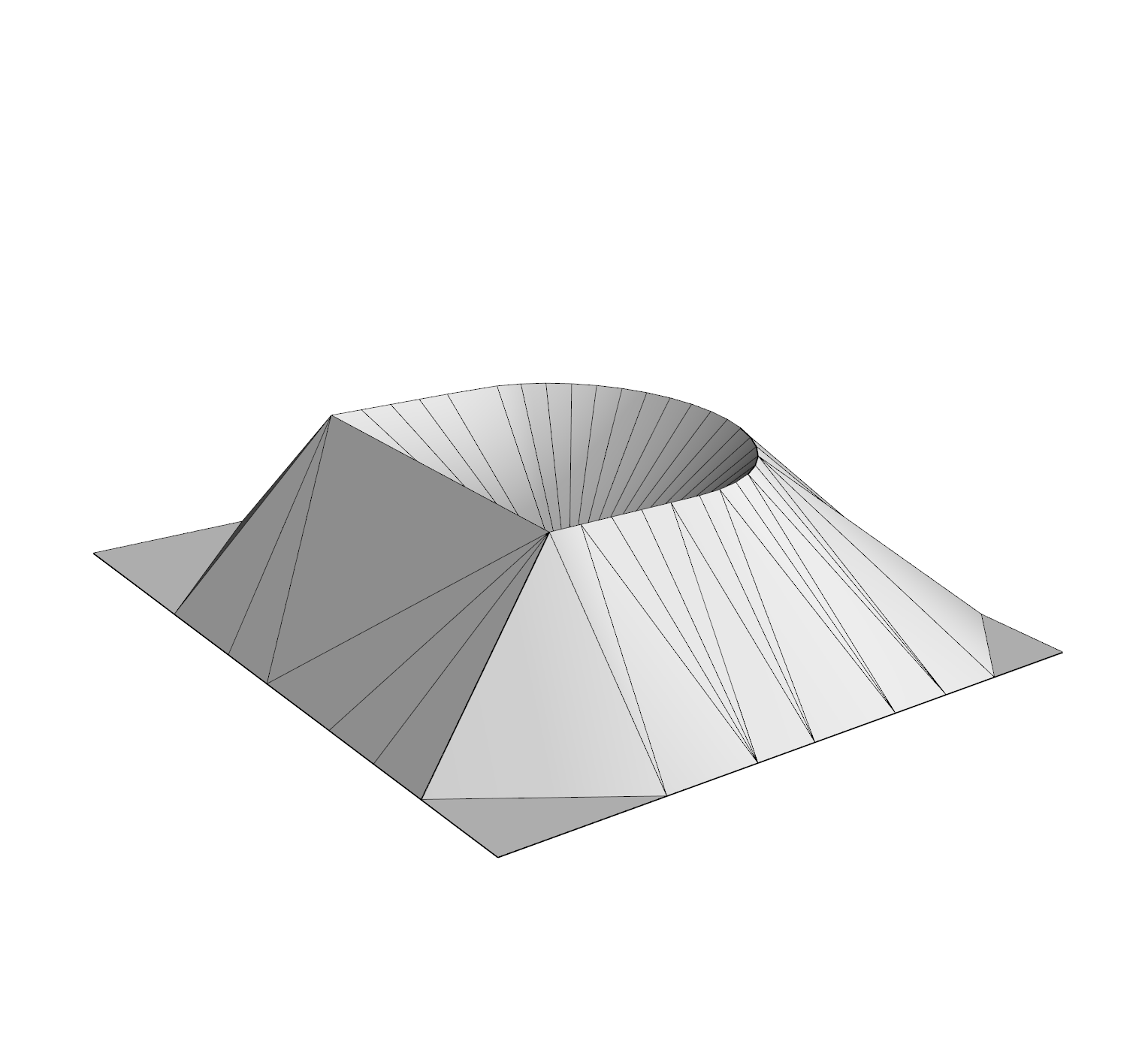}
    \caption{A function and its linear extension onto the Delaunay triangulation of a sample.  The discretization introduces a spurious feature because it does not register the opening.}\label{fig:horseshoe}
\end{figure}

The most direct way to use a Delaunay triangulation to estimate the persistence diagram of a set of points is to extend the function from the vertex set to the simplices, setting the value at the simplex be the maximum value at its vertices.
This corresponds to the piecewise linear extension of the function to the geometric realization of the complex (see Section 2.5, Morozov~\cite{morozov08thesis}).
Unfortunately, this can fail badly in the absence of strong sampling guarantees;
Figure~\ref{fig:horseshoe} shows an example where the piecewise linear extension introduces spurious features.
Thus, the problem with some samples is not only that we can miss features, but that we may hallucinate them as well.

The key to using a discretization is to use an image filtration (see Cohen-Steiner et al.~\cite{cohen-steiner09persistent}, Bauer~\cite{bauer22efficient}) capturing both an upper and lower bound.
The following theorem states that we can compute a sub-barcode for any Lipschitz function by filtering the Delaunay triangulation by upper and lower bounds given on the input values and the radius of the Voronoi cells.

\begin{theorem}\label{thm:delaunay_filtration}
    Let $S$ be a finite subset of a polytope $X\subset\R^d$.
    Let $f_S:S\to \RR$ be a sample of an unknown Lipschitz function $f:X\to \RR$.
    For each $s\in S$, let $r_s$ denote the radius of the Voronoi cell $\vor_S(s)$.
    Let $u$ and $\ell$ be functions $\del_S\to\RR$ defined on Delaunay simplices $\sigma\in\del_S$ as
    \[
        u(\sigma):= \max_{s\in \sigma}f_S(s) + r_s \text{~~and~~} \ell(\sigma):= \max_{s\in \sigma}f_S(s) - r_s.
    \]
    Then, $\bar(u\ge l)\subb \bar(f)$.
\end{theorem}

\begin{proof}
    To relate the barcodes, we will first construct functions on $X$ that have the same barcode as the simplicial functions $u$ and $\ell$. 
    Second, we will relate these functions to the Lipschitz extensions $\check{f}_S$ and $\hat{f}_S$.

    We define piecewise constant functions on $X$ using the Voronoi cells and $f_S$ as
    \[
        u^*(x):= \max_{s\mid x\in \vor_S(s)}f_S(s) + r_s \text{~~and~~} \ell^*(x):= \max_{s\mid x\in \vor_S(s)}f_S(s) - r_s.
    \]
    By the Persistent Nerve Lemma (Chazal et al.~\cite{chazal08towards}, Lemma 3.4), there is a homotopy equivalence between the sublevels of $u$ and $u^*$ that commutes with inclusions (see also \cite{bauer23unified}). 
    The same holds for $\ell$ and $\ell^*$.
    It follows that $\bar(u\ge \ell) \cong \bar(u^*\ge \ell^*)$.

    For any $x\in X$, there is an $s\in S$ such that $x\in \vor_S(s)$ and $u^*(x) = f_S(s) + r_s$, so
    \[
        u^*(x) = f_S(s) + r_s \ge f_S(s) + \dist(x,s) \ge \min_{s_0\in S} f_S(s_0) + \dist(x,s_0) = \check{f}_S(x).
    \]
    Similarly, there is an $s\in S$ such that $x\in \vor_S(s)$ and $\ell^*(x) = f_S(s) - r_s$, so
    \[
        \ell^*(x) = f_S(s) - r_s \le f_S(s) - \dist(x,s) \le \max_{s_0\in S} f_S(s_0) - \dist(x,s_0) = \hat{f}_S(x),
    \]
    thus $u^* \ge \check{f}_S\ge \hat{f}_S\ge \ell^*$, and Corollary~\ref{cor:subbarcodes} therefore implies
    \[
        \bar(u\ge \ell) \cong \bar(u^*\ge \ell^*) \subb \bar(\check{f}_S\ge \hat{f}_S) \subb \bar(f).\qedhere
    \]
\end{proof}

\subsection{Semi-Supervised TDA}

We have assumed thus far that we have access to function values at all points in the sample.
This resembles a supervised learning problem.
If instead we only have function values at a subset $P\subset S$, we can still use the points of $S$ to improve our approximation.
We call this \emph{semi-supervised TDA}.

The most common guarantees in PH are derived from interleavings, and thus yield bottleneck distance bounds on the resulting barcodes.
In this section, we combine those results with the Lipschitz extension sub-barcode and the Delaunay filtration above to get a guaranteed approximation.
Throughout, we assume that $S$ is a sample of $X$, and that we only have function values $f_P:P\to \R$ for $P\subset S$.
The algorithm extends $\check{f}_P$ and $\hat{f}_P$ to the points of $S$ and then use the Delaunay filtration from the previous subsection.

In the following, we say that $S$ is an \emph{$\e$-sample} of $X$ if every point of $X$ is within $\e$ of a point in $S$.
Equivalently, the radius of every Voronoi cell $\vor_S(s)$ is at most $\e$.

\begin{theorem}\label{thm:semi_supervised}
    Let $X\subset\R^d$ be a convex polytope, and suppose $S\subset X$ is an $\e$-sample.
    Let $f_P:P\to \RR$ be a sample of an unknown Lipschitz function $f : X\to\RR$ defined on a subset $P\subset S$,
    and let $u$ and $\ell$ be functions $\del_S\to \RR$ defined on each Delaunay simplex $\sigma \subseteq S$ as
    \[
        u(\sigma) = \max_{s\in \sigma}\check{f}_P(s) + \e \text{~~and~~} \ell(\sigma) = \max_{s\in \sigma}\hat{f}_P(s) - \e.
    \]
    Then, $\bar(u\ge \ell)\subb \bar(f)$, and
        $\dist_B\big(\bar(u\ge \ell), \bar(\check{f}_P\ge \hat{f}_P)\big) \le \e$.
\end{theorem}

\begin{proof}
    The sub-barcode relation follows from Theorem~\ref{thm:delaunay_filtration} and the observation that the radius $r_s\le \e$ for all $s$ in an $\e$-sample.
    The bottleneck bound follows from a more general theory of interleaving of image persitence modules.
    It depends only on the fact that the extensions $u^*$ and $\ell^*$ of $u$ and $\ell$ to piecewise constant functions on Voronoi cells 
    (as in the proof of Theorem~\ref{thm:delaunay_filtration}) satisfy $\|u^* - \check{f}_P\|_{\infty}\le \e$ and $\|\ell^* - \hat{f}_P\|_{\infty}\le \e$, respectively.
\ARXIV{
    These bounds yeild an $\e$-interleaving of the sublevels, and thus give an interleaving of the images.\footnote{See Theorem~\ref{thm:image_stability} for the more general theorem for stabilty of image persistence.}
\end{proof}
}
\SOCG{
    These bounds yeild an $\e$-interleaving of the sublevels, and thus give an interleaving of the images.\footnote{A more general theorem for stabilty of image persistence can be found in the full version of this paper~\cite{chubet22theory}.}
\end{proof}
}

\begin{figure}[!ht]
    \centering
    \includegraphics[trim=200 150 200 150, clip, width=0.3\textwidth]{\SURF/\SAMPLE/lips/delaunay/\SAMPLE_delaunay-im-lips-max-color-\SUBN subsample550}
    \includegraphics[trim=200 150 200 150, clip, width=0.3\textwidth]{\SURF/\SAMPLE/lips/delaunay/\SAMPLE_delaunay-im-lips-max-color-\SUBN subsample717}
    \includegraphics[trim=200 150 200 150, clip, width=0.3\textwidth]{\SURF/\SAMPLE/lips/delaunay/\SAMPLE_delaunay-im-lips-max-color-\SUBN subsample1017}\\
    \includegraphics[trim=200 150 200 150, clip, width=0.3\textwidth]{\SURF/\SAMPLE/lips/delaunay/\SAMPLE_delaunay-im-lips-min-color-\SUBN subsample550}
    \includegraphics[trim=200 150 200 150, clip, width=0.3\textwidth]{\SURF/\SAMPLE/lips/delaunay/\SAMPLE_delaunay-im-lips-min-color-\SUBN subsample717}
    \includegraphics[trim=200 150 200 150, clip, width=0.3\textwidth]{\SURF/\SAMPLE/lips/delaunay/\SAMPLE_delaunay-im-lips-min-color-\SUBN subsample1017}\\
    \includegraphics[width=\textwidth]{\SURF/\SAMPLE/lips/delaunay/\SAMPLE_barcode-lips-sub\SUBN}
    \caption{The top row shows the Delaunay filtration of the max Lipschitz extension. The next row shows the min extension. 
        The barcode of the image suppresses the small spurious features and captures the two main features.}\label{fig:delaunay_lips_sub}
\end{figure}

\subsection{Finer Approximation with Barycentric Subdivision}

In general, we would like to have tight approximations to the sub-barcode of Lipschitz extensions without requiring any sampling assumptions.
One way to get a better bound is to manually add points to reduce the radii of the cells.
Another approach would be to consider the barycentric subdivision.
This makes sense because it can be understood as assigning function values to Delaunay simplices individually rather than inducing the values from the function on the vertices.
In this section, we show how to define such constructions in a way that applies to a much more general class of functions.

For each sample $s\in S$, knowing the value of $f(s)$ implies an upper bound and a lower bound on $f$.
Thus far, these bounds came from the assumption that $f$ is Lipschitz.
Similar upper and lower bounds can be defined for other classes of functions.
We only assume that these bounds are \emph{distance monotone}.
That is, the upper bound $\check{f}_s$ for a point $s$ satisfies the condition that if $\dist(x,s)\le \dist(y,s)$, then $\check{f}_s(x)\le \check{f}_s(y)$.
For the lower bound $\hat{f}_s$ at $s$ values decrease with distance, i.e., if $\dist(x,s)\le \dist(y,s)$, then $\hat{f}_s(x)\ge \hat{f}_s(y)$.
We call $\{\check{f}_s\}$ and $\{\hat{f}_s\}$ a family of \emph{pointwise distance monotone bounds}.
As before, we let $\check{f}_S(x):= \min_{s\in S}\check{f}_s(x)$ and $\hat{f}_S(x):= \max_{s\in S}\hat{f}_s(x)$.

Although the term baryentric subdivision implies that subdivision happen at the barycenters of the cells, we define this operation instead as a combinatorial operation on simplicial complexes.
For any simplicial complex $K$, we define $\bary K$ to be the simplicial complex with a vertex for each simplex in $K$ and  a simplex for each ordered subset of simplices (the ordering is by inclusion).
For any function $g:K\to \RR$, we can extend it to a filtration $G$ on $\bary K$, as $\sigma_0\into \cdots \into \sigma_k$ is in $G(t)$ if and only if $g(\sigma_i)\le t$ for all $i\in \{0,\ldots, k\}$.
\SOCG{
    Proof of the following theorem may be found in the full version~\cite{chubet22theory}.
}
\ARXIV{
    Proof of the following theorem may be found in Appendix~\ref{sec:proofs}.
}

\begin{theorem}\label{thm:bary}
    Let $S$ be a finite sample of a convex polytope $X$, 
    and let $\{\check{f}_s:\vor_S(s)\to \RR\mid s\in S\}$ and $\{\hat{f}_s:\vor_S(s)\to \RR\mid s\in S\}$ be families of pointwise distance-monotone bounds on some unknown function $f : X\to \RR$.
    Aggregate these bounds as $\check{f}_S := \Meet_{s\in S}\check{f}_s$ and $\hat{f}_S:=\Join_{s\in S}\hat{f}_s$.
    For $\sigma \in \del_S$, let 
    \[
        u(\sigma) := \min_{x\in \vor_S(\sigma)}\check{f}_S(x) \text{~~and~~} \ell(\sigma) := \min_{x\in \vor_S(\sigma)}\hat{f}_S(x).
    \]
    Extend $u$ and $\ell$ to $\bary \del_S$.
    Then, $\bar(u\ge \ell)= \bar(\check{f}_S\ge \hat{f}_S) \subb \bar(f)$.
\end{theorem}

\begin{example}\label{ex:alpha}
    To apply this theorem to the special case of Lipschitz functions, the resulting filtration captures the usual Lipschitz upper and lower bounds restricted to the Voronoi cells.
    That is, 
    \[
        \check{f}_s:\vor_S(s)\to \RR : x\mapsto f_S(s) + \dist(x,s)
    \text{~~and~~}
        \hat{f}_s:\vor_S(s)\to \RR : x\mapsto f_S(s) - \dist(x,s).
    \]

    Letting $\alpha(\sigma):= \min_{x\in \vor_S(\sigma)}\dist(x,\sigma)$ be the birth time of the simplex $\sigma$ in the standard Delaunay filtration (also known as the $\alpha$-complex filtration).
    Then, the corresponding filtration on $\bary \del_S$ is induced by
    \[
        u(\sigma):= \max_{s\in \sigma}f_S(s) + \alpha(\sigma),
    \text{~~and~~}
        \ell(\sigma):= \min_{s\in \sigma}f_S(s) - \alpha(\sigma).
    \]
\end{example}

\section{Categorical Constructions}\label{sec:categorical}

The following section describes how the concepts presented above fit neatly into a category-theoretic framework.
It contains some technical concepts that may unfamiliar to computational geometers.

\subsection{Smoothing Persistence Modules}

Given a barcode $B$, an $\e$-smoothing is a new barcode defined by eliminating bars of length at most $2\e$, and replacing all other bars $[b,d]$ with a $[b+\e, d-\e]$;
that is, we trim $\e$ from the beginning and the end of each bar.
The idea of a smoothed barcode naturally arises in the literature on the stability of persistence modules~\cite{chazal09proximity,bubenik15metrics,lesnick15theory}.
It is clear from the definition that the smoothed barcode is a sub-barcode of the original.
This relationship can be expressed clearly as arising from a factorization as follows.

Let $\delta_{-}$ and $\delta_{+}$ be order-preserving maps $\R\to \R$ with the property that, for all $t\in \R$, we have $\delta_{-}(t)\le t\leq \delta_+(t)$.
Order preserving maps between posets are functors, so the pointwise order relations on $\delta_{-}$ and $\delta_{+}$ correspond to natural transformations
\[
    \e_0: \delta_{-}\Rightarrow \id_\RR \text{~~~and~~~} \e_1: \id_\RR\Rightarrow \delta_{+}.
\]
So, for any persistence module $M:\RR\to \vec$, composition with $M$ yields the following factorization.

\begin{equation}\label{dgm:factorization}
    \centering
    \begin{tikzcd}
      M\e_0\arrow[rr,""]\arrow[dr,""'] && M\e_1\\
      & \id_M\arrow[ur,""']
    \end{tikzcd}
\end{equation}
Letting $\e = \e_1\circ\e_0$, we can view $M\e$ (or equivalently, its image) as the smoothed persistence module.
The factorization induces the sub-barcode matching in the obvious way:
\[
    \bar(M\e)\subb \bar(M).
\]
From this perspective, an interleaving of persistence modules may be viewed as a pair of compatible factorizations rather than a pair of compatible (shifted) homomorphisms.
In fact, the pair $(\delta_{-},\delta_+)$ corresponds to an \emph{adjunction} with counit $\e_0$ and unit $\e_1$ (see~\cite{gardner22thesis}).

\subsection{Sub-Barcodes are Subobjects}

Within the literature, there are several different categorifications of persistence barcodes (or equivalently persistence diagrams).
Here, we identify a natural way to define a category of barcodes that plays nicely with our treatment of sub-barcodes.


First, recall that unlike much of the prior work, we do not regard barcodes as multisets.
Instead, we define a barcode to be a function from a set to the intervals.
In computer programming terms, we have an ``is a'' versus ``has a'' distinction;
in the multiset view, each bar \emph{is an} interval.
In our view, each bar \emph{has an} interval.

A standard categorical approach to construct a category $\int_\cat{C} F$ from a (pseudo)functor $\cat{C}\to \Cat$ is the Grothendieck construction.
Let $F : \op{\cat{C}}\to \Cat$ be a functor from a category \cat{C} to the category \Cat of (small) categories.
For the purposes of this work, the \emph{(covariant) Grothendieck construction} yields a category $\int_\cat{C} F$ with objects given by pairs $(C,x)$ for each $C\in\cat{C}$ and $x\in F(C)$,
and arrows $(f,h) : (C,x)\to (C',x')$ for each pair of arrows $f : C\to C'$ in $\cat{C}$ and $h : x \to F\b{f}(x')$ in $F(C)$.

Let $[-, \I] : \Set \to \Cat$ be the (pseudo)functor that takes a set $X$ and returns the category $\I^X = [X,\I]$ of functors $X\to \I$.
Then the category \Bar of barcodes described in Section~\ref{sec:subbarcodes} is a contravariant Grothendieck construction
\[
    \Bar = \int_{\Set}[-,\I].
\]
The objects of \Bar are pairs $(\ul{B}, B:\ul{B}\to \I)$---i.e. barcodes---and a morphism of \Bar $(\ul{A},A) \to (\ul{B},B)$ is a function $\phi:\ul{A}\to \ul{B}$ such that for all $\alpha\in \ul{A}$, we have that $A(\alpha)\subseteq B(\phi(\alpha))$.
To see this as a Grothendieck construction, it suffices to observe that the barwise inclusion ordering is a natural transformation $A\Rightarrow B\phi$ in $[\ul{A},\I]$.

The condition on morphisms ensures that a bar $\alpha$ can only map to a bar $\beta$ when $A(\alpha)\subseteq B(\beta)$.
Thus, a injective morphism is exactly a sub-barcode matching.
These injective morphisms are the monomorphsisms in this category and thus sub-barcodes are subobjects in this category.

\begin{remark}
    The category \Bar defined above is also known as the category of $\I$-fuzzy sets~\cite{goguen67fuzzy},
    and the construction of a category of fuzzy sets using the Grothendieck construction also noted by Jardine~\cite{jardine2019fuzzy}.
    It is also possible to replace $\Set$ with $\Mch$, the category of sets with partial injective maps (i.e., matchings), to construct barcodes as $\int_{\Mch}[-,\I]$,
    or more generally, to replace $\Set$ with the category $\Set_+\supset \Mch$ of sets and \emph{partial functions}.
    In both cases, the monomorphisms are injective sub-barcode matchings that yield sub-barcode relations.
\end{remark}

\subsection{Ranks via Presheaves}


The rank functor was previously constructed directly from a persistence module, or more generally, a factorization of persistence module homomorphisms.
In this section, we show how the rank functor can also be constructed from a barcode.
This is already well-known; the novelty here is that we show how the rank functor can be factored through the category of barcodes.
This gives the more abstract proof that sub-barcodes are more discriminating than ranks.

For any category $\cat{C}$, there is a functor $L:\int_{\Set}[-,\cat{C}]\to [\op{\cat{C}}, \Set]$ given for $A:\underline{A}\to \cat{C}$ by
\[
    LA(c) = \{a\in \underline{A} \mid \exists c\to A(a)\}.
\]
The morphisms $L(A)(c\to c')$ are inclusions $L(A)(c')\subseteq L(A)(c)$.
For $(f,\e):A\to B$ in $\int_{\Set}[-,\cat{C}]$, it is easy to check that $f$ gives a natural transformation $LA\Rightarrow LB$.

For the special case where $\cat{C} = \I$, this is a functor from barcodes to presheaves of intervals.
The intuitive meaning is that, for a barcode $B$, the presheaf $LB$ maps an interval $J$ to the set of bars in $B$ that contain all of $J$.
All morphisms in the image of $L$ are inclusions, and so, for $J\supseteq I$, we have $LB(J)\subseteq LB(I)$ and thus $|LB(J)|\le |LB(I)|$.
Here the vertical bars indicate set cardinality which is a functor $\card$ from the poset of finite sets with inclusions to the poset $\NN$.
Thus, we have a functor $\card\circ L:\Bar\to \Set^{\op{\I}}$.



It is now an easy exercise to show that $\rank = \card\circ L\circ \bar$.
All morphisms in the functor category $[\I^{op}, \NN] = \NN^{\op{\I}}$ are monomorphisms, so clearly a sub-barcode relation (a monomorphism of barcodes) yields a subrank relation (a monomorphism of rank invariants).
The other direction doesn't hold.
The most natural way to define a barcode from the rank invariant does not give a functor to barcodes.
The example is given in the proof of Theorem~\ref{thm:subbarcodes_vs_ranks}.

\begin{remark}
    Barr~\cite{barr86fuzzy} showed that, when the indexing poset $L$ is a Heyting algebra, the category of $L$-fuzzy sets is equivalent to a category of sheaves of monomorphisms.
    Following this result, many of the recent applications of fuzzy sets~\cite{spivak09metric,mcinnes18umap} have defined fuzzy sets directly as functors $L\to \Set$.
    Unfortunately, because the poset of intervals $\I$ is not a Heyting algebra, this equivalence cannot be applied directly to the category $\Bar$.
    On the other hand, the category of intervals with the product or Egli-Milner ordering is a Heyting algebra.
    Extending Barr's equivalence to a category of barcodes and partial functions is the subject of future work.
\end{remark}

\bibliographystyle{plainurl}
\bibliography{bibliography}

\begin{thebibliography}{10}

\bibitem{barr86fuzzy}
Michael Barr.
\newblock Fuzzy set theory and topos theory.
\newblock {\em Canadian Mathematical Bulletin}, 29(4):501--508, 1986.

\bibitem{bauer23unified}
Ulrich Bauer, Michael Kerber, Fabian Roll, and Alexander Rolle.
\newblock A unified view on the functorial nerve theorem and its variations.
\newblock {\em Expositiones Mathematicae}, 41(4):125503, 2023.
\newblock URL:
  \url{https://www.sciencedirect.com/science/article/pii/S0723086923000415},
  \href {https://doi.org/https://doi.org/10.1016/j.exmath.2023.04.005}
  {\path{doi:https://doi.org/10.1016/j.exmath.2023.04.005}}.

\bibitem{bauer13induced}
Ulrich Bauer and Michael Lesnick.
\newblock Induced matchings and the algebraic stability of persistence
  barcodes.
\newblock {\em Journal of Computational Geometry}, page Vol. 6 No. 2 (2015):
  Special issue of Selected Papers from SoCG 2014, 2015.
\newblock URL: \url{https://jocg.org/index.php/jocg/article/view/2983}, \href
  {https://doi.org/10.20382/JOCG.V6I2A9} {\path{doi:10.20382/JOCG.V6I2A9}}.

\bibitem{bauer22efficient}
Ulrich Bauer and Maximilian Schmahl.
\newblock Efficient computation of image persistence.
\newblock {\em arXiv preprint arXiv:2201.04170}, 2022.

\bibitem{bubenik15metrics}
Peter Bubenik, Vin De~Silva, and Jonathan Scott.
\newblock Metrics for generalized persistence modules.
\newblock {\em Foundations of Computational Mathematics}, 15(6):1501--1531,
  2015.

\bibitem{buchet15topological}
Micka{\"e}l Buchet, Fr{\'e}d{\'e}ric Chazal, Tamal~K. Dey, Fengtao Fan,
  Steve~Y. Oudot, and Yusu Wang.
\newblock Topological analysis of scalar fields with outliers.
\newblock In {\em 31st International Symposium on Computational Geometry (SoCG
  2015)}, volume~34 of {\em Leibniz International Proceedings in Informatics
  (LIPIcs)}, pages 827--841. Schloss Dagstuhl--Leibniz-Zentrum fuer Informatik,
  2015.

\bibitem{chazal09proximity}
Fr{\'e}d{\'e}ric Chazal, David Cohen-Steiner, Marc Glisse, Leonidas~J. Guibas,
  and Steve~Y. Oudot.
\newblock Proximity of persistence modules and their diagrams.
\newblock In {\em Proceedings of the Twenty-fifth Annual Symposium on
  Computational Geometry}, SCG '09, pages 237--246, New York, NY, USA, 2009.
  ACM.
\newblock URL: \url{http://doi.acm.org/10.1145/1542362.1542407}, \href
  {https://doi.org/10.1145/1542362.1542407}
  {\path{doi:10.1145/1542362.1542407}}.

\bibitem{chazal11scalar}
Fr{\'e}d{\'e}ric Chazal, Leonidas~J. Guibas, Steve~Y. Oudot, and Primoz Skraba.
\newblock Scalar field analysis over point cloud data.
\newblock {\em Discrete {\&} Computational Geometry}, 46(4):743--775, 2011.

\bibitem{chazal08towards}
Fr{\'e}d{\'e}ric Chazal and Steve~Yann Oudot.
\newblock Towards persistence-based reconstruction in euclidean spaces.
\newblock In {\em Proceedings of the Twenty-fourth Annual Symposium on
  Computational Geometry}, SCG '08, pages 232--241, New York, NY, USA, 2008.
  ACM.
\newblock URL: \url{http://doi.acm.org/10.1145/1377676.1377719}, \href
  {https://doi.org/10.1145/1377676.1377719}
  {\path{doi:10.1145/1377676.1377719}}.

\bibitem{chubet22theory}
Oliver~A. Chubet, Kirk~P. Gardner, and Donald~R. Sheehy.
\newblock A theory of sub-barcodes, 2022.
\newblock URL: \url{https://arxiv.org/abs/2206.10504}, \href
  {https://doi.org/10.48550/ARXIV.2206.10504}
  {\path{doi:10.48550/ARXIV.2206.10504}}.

\bibitem{cohen2010lipschitz}
David Cohen-Steiner, Herbert Edelsbrunner, John Harer, and Yuriy Mileyko.
\newblock Lipschitz functions have l p-stable persistence.
\newblock {\em Foundations of computational mathematics}, 10(2):127--139, 2010.

\bibitem{cohen-steiner09persistent}
David Cohen-Steiner, Herbert Edelsbrunner, John Harer, and Dmitriy Morozov.
\newblock Persistent homology for kernels, images, and cokernels.
\newblock In {\em SODA: ACM-SIAM Symposium on Discrete Algorithms}, 2009.

\bibitem{edelsbrunner02topological}
Edelsbrunner, Letscher, and Zomorodian.
\newblock Topological persistence and simplification.
\newblock {\em Discrete {\&} Computational Geometry}, 28(4):511--533, Nov 2002.
\newblock \href {https://doi.org/10.1007/s00454-002-2885-2}
  {\path{doi:10.1007/s00454-002-2885-2}}.

\bibitem{gardner22thesis}
Kirk~Patrick Gardner.
\newblock {\em Verified Topological Data Analysis and a Theory of
  Sub-Barcodes}.
\newblock PhD thesis, North Carolina State University, 2022.

\bibitem{goguen67fuzzy}
Joseph~A Goguen.
\newblock L-fuzzy sets.
\newblock {\em Journal of mathematical analysis and applications},
  18(1):145--174, 1967.

\bibitem{jardine2019fuzzy}
John~F. Jardine.
\newblock Fuzzy sets and presheaves.
\newblock {\em Compositionality}, 1:3, dec 2019.
\newblock URL: \url{https://doi.org/10.32408%2Fcompositionality-1-3}, \href
  {https://doi.org/10.32408/compositionality-1-3}
  {\path{doi:10.32408/compositionality-1-3}}.

\bibitem{lawvere73metric}
F~William Lawvere.
\newblock Metric spaces, generalized logic, and closed categories.
\newblock {\em Rendiconti del seminario mat{\'e}matico e fisico di Milano},
  43:135--166, 1973.

\bibitem{lesnick15theory}
Michael Lesnick.
\newblock The theory of the interleaving distance on multidimensional
  persistence modules.
\newblock {\em Foundations of Computational Mathematics}, 15(3):613--650, 2015.

\bibitem{mcinnes18umap}
Leland McInnes, John Healy, and James Melville.
\newblock Umap: Uniform manifold approximation and projection for dimension
  reduction.
\newblock {\em arXiv preprint arXiv:1802.03426}, 2018.

\bibitem{morozov08thesis}
Dmitriy Morozov.
\newblock {\em Homological Illusions of Persistence and Stability}.
\newblock PhD thesis, Duke University, 2008.

\bibitem{spivak09metric}
David~I Spivak.
\newblock Metric realization of fuzzy simplicial sets.
\newblock Self-published Notes.

\end{thebibliography}

\ARXIV{
    \appendix
    \section{Omitted Proofs}\label{sec:proofs}

\begin{proof}[Proof of Theorem~\ref{thm:bary}]
    For each $t\in \RR$, we define a deformation retraction from the sublevels of the bounds to the corresponding subcomplex of the barycentric subdivision.
    The deformation retraction implies the desired equivalence of barcodes.

    The function $u$ is monotone on simplices in the sense that if $\sigma \subset \tau$ then $u(\sigma)\ge u(\tau)$.
    This follows from the monotonicity of the bounds.
    So, $u$ could give a filtration on $\del_S$ without subdivision.
    However, $\ell$ is a cofiltration, i.e., it is reversed with respect to simplex ordering.
    Thus, the barycentric subdivision allows both filtrations to be represented on the same complex.
    Both filtrations are induced filtrations (defined only by the values at the vertices).

    The deformation retractions from the sublevels of $\check{f}_S$ and $\hat{f}_S$ are defined by affine projection.
    This projection is easiest to express in the geometric realization of the complex where vertex is treated as a basis vector.
    Then, every point in the complex is a convex combination of vertices.
    Let $V$ be the vertices $v$ such that $\check{f}_S(v)\le t$.
    The affine projection of the $t$-sublevels of $\check{f}_S$ is defined as
    \[
        p(x) := \frac{\sum_{v\in V} x_v v}{\sum_{v\in V}x_v}.
    \]
    Every point is fixed unless it lies in a simplex for which some vertices have value at most that $t$ and other have value greater than $t$.

    When forming $\bary\del_S$, we have options about where to place the vertices associated with simplices.
    We embed the vertex for $\sigma$ at the point of $\vor_S(\sigma)$ that is closest to $S$.
    If multiple simplices map to the same point, then it is a simple matter to collapse them (and the corresponding simplices) to a single vertex.
    As all function values are determined by distance to $S$, these points have the same value and do not affect the computation.

    Let $x$ be a point in a geometric simplex $\sigma$.
    If we partition the vertices of $\sigma$ into sets $N$ and $F$ (think {\ul N}ear and {\ul F}ar) so that all the vertices of $N$ are nearer to $S$ than every vertex of $F$.
    Then, there is an affine projection $p$ of $x$ on $N$ and another projection $q$ of $x$ onto $F$.
    It is  a straightforward geometric computation to see that this projection is distance monotone (see Lemma~\ref{lem:distance_monotone_projection} for the complete proof of this fact).
    So, projecting onto the sublevels of $u$ gives a deformation retraction, because the fibers of the projection are all simply connected (star-shaped).
    The same holds for the affine projection to the sublevels of $\ell$.

    A deformation retraction of each sublevel set induces an equivalence of persistence modules and so $\bar(u\ge \ell) = \bar(\check{f}_S \ge \hat{f}_S)$.
\end{proof}


\section{Image Stability}\label{sec:image_stability}

In this section, we prove a basic lemma about the stability of image persistence modules.

\begin{theorem}\label{thm:image_stability}
    Let $\e\ge \id_{\RR}$ be a shift in $\RR$, and let $(F,G):A\to B$ and $(H,K):C\to D$ be $\e$-interleavings.
    If $\phi:A\to D$ and $\psi:B\to C$ commute with the interleavings in the sense that
    \[
        K\phi = (\psi\e)F \text{~~~and~~~} H\psi = (\phi\e)G,
    \]
    Then $\dist_I(\im \phi, \im \psi)\le \e$.
\end{theorem}

\begin{proof}
    The \emph{arrow category} associated with the category $\vec^\RR$ of persistence modules is the functor category $[\cat{2}, \vec^\RR]\simeq [\RR\times \cat{2},\vec]$, where $\cat{2} = \{0\leq 1\}$ is the poset of truth values.
    The commutativity conditions of the interleaving imply that, in the arrow category of persistence modules, we have $(F,K):\phi\to\psi\e$ and $(G,H):\psi\to \phi\e$.
    These morphisms give an $\e$-interleaving of $\phi$ and $\psi$.
    The universal property of images then gives the corresponding interleaving of their images. 
\end{proof}

    \section{Distance Monotone Projections}\label{sec:apx_projection}

\begin{figure}
    \includegraphics{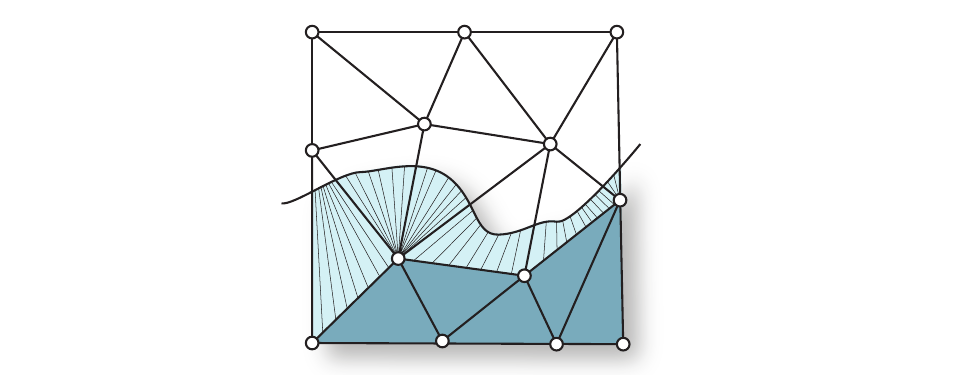}
    \caption{An illustration of affine projection of a sublevel set onto a subcomplex.}\label{fig:affine_projection}
\end{figure}

When subdividing the Voronoi diagram (equivalently the Delaunay triangulation), we use a \emph{minimum distance embedding}.
That is, we embed $\sigma_i$ at the point $v_i$ of $\vor_S(\sigma)$ that is closest to $S$.
In this embedding, the vertices are naturally ordered by distance.
That is, an ordered sequence of simplices $\sigma_0\into \cdots \into \sigma_k$, naturally has the property that their embedding satisfies
\[
    \dist(v_0, S) \le \cdots \le \dist(v_k, S)
\]
In fact, an even stronger fact is true, for all $i< j$ and all $s\in \sigma_i$, we have
\[
    (v_i-s)^\top(v_j-v_i)\ge 0.
\]
This not only implies that $\|v_i-s\|\le \|v_j-s\|$, but also that the entire line segment between the two monotone with respect to distance.

For a simplex of the subdivision $\sigma_0\into \cdots \sigma_k$, consider a partition into near and far vertices, i.e., $N = \{v_0\ldots ,v_a\}$ and $F = \{v_{a+1},\ldots,v_k\}$.
Let $p$ be a point in the subsimplex spanned by $n$ and let $q$ be a point in the subsimplex spanned by $F$.
In coordinates, we write these as convex combinations:
\[
    p = \sum_{v_i\in N}p_iv_i \text{~~and~~} q = \sum_{v_j\in F} q_jv_j.
\]

The following lemma shows that distance increases monotonely along the straight line from $p$ to $q$.

\begin{lemma}\label{lem:distance_monotone_projection}
    For all $p$ and $q$ constructed as above, we have $(p-s)^{\top}(q-p)\ge 0$ for all $s$ such that $p$ and $q$ in $\vor_S(s)$.
\end{lemma}
\begin{proof}
    Without loss of generality, we assume that $s = 0$.
    First, we observe that for all $v_i\in N$, we have
    \begin{align*}
        v_i^\top(q - v_i)
            &= v_i^\top\left(\left(\sum_{v_j\in F}q_j v_j\right) - v_i\right)\\
            &= v_i^\top\left(\sum_{v_j\in F}q_j (v_j-v_i)\right)\\
            &= \sum_{v_j\in F}q_j v_i^\top(v_j-v_i)\\
            &\ge 0.
    \end{align*}
    The last line follows because if $v_i\in N$ and $v_j\in F$, then $i< j$ and so $v_i^\top(v_j-v_i)\ge 0$.
    
    By the triangle inequality and the Cauchy-Schwarz inequality, we have
    \begin{align*}
        p^\top p
            &= \|p\|^2\\
            &\le \left(\sum_{v_i\in N}p_i\|v_i\|\right)^2\\
            &\le \left(\sum_{v_i\in N}(\sqrt{p_i})^2\right)\left(\sum_{v_i\in N}(\sqrt{p_i}\|v_i\|)^2\right)\\
            &= \sum_{v_i\in N}p_i v_i^\top v_i.
    \end{align*}

    So, we complete the proof that $p^\top(q-p)\ge 0$ as follows.
    \begin{align*}
        p^\top(q-p)
            &= p^\top q - p^\top p\\
            &\ge \left(\sum_{v_i\in N}p_iv_i^\top q\right) - \left(\sum_{v_i\in N} p_i v_i^\top v_i\right)\\
            &= \sum_{v_i\in N} p_i v_i^\top (q-v_i)\\
            &\ge 0.
    \end{align*}
    The last line follows because we have already shown that each of the terms in the sub is nonnegative. 
\end{proof}

}

\end{document}